\documentclass[
  prl,
  twocolumn,
  citeautoscript,
  showpacs,
  amsart,
  longbibliography
]{revtex4-2}

\usepackage{amsmath,bm,amsfonts}
\usepackage{physics,mathtools}
\usepackage{graphicx}
\usepackage[normalem]{ulem}
\usepackage{titlesec}
\usepackage[colorlinks=true,linkcolor=blue,citecolor=blue,urlcolor=blue]{hyperref}

\usepackage[dvipsnames]{xcolor}

\graphicspath{{figures/}}

\date{\today}

\begin{document}

\title{Roughening Transition in Quantum Circuits}

\author{Hyunsoo Ha}
\author{David A. Huse}
\author{Grace M. Sommers}
\affiliation{Department of Physics, Princeton University, Princeton, NJ 08544, USA}

\begin{widetext}
\begin{abstract}
We explore a roughening phase transition that occurs in the entanglement dynamics of certain quantum circuits.
Viewing entanglement as the free energy of a membrane in a circuit-defined random environment, there is a competition between membrane smoothing due to lattice pinning and roughening due to disorder in the circuit.  In particular, we investigate the randomness-induced roughening transition of the entanglement membrane in a (3+1)-dimensional Clifford circuit model, by calculating the entanglement entropy for various bipartitions.  We further construct a scaling theory for membranes tilted away from lattice planes, uncovering new scaling forms and a crossover to a previously unexplored critical “tilted regime”.
\end{abstract}
\end{widetext}

\maketitle

The dynamics of entanglement in quantum many-body systems is a central problem~\cite{Amico_Vedral_RMP2008}, connecting fundamental phenomena such as thermalization and quantum chaos~\cite{page_prl1993,Kim_Huse_2013,d'alessio_rigol_review2016,kaufman_science2016,bianchi_vidmar_prxq2022} with dynamical phases of matter~\cite{nandkishore_huse_mbl,gopalakrishnan_huse_prb2019,alet_laflorencie_mbl_review,Abanin_Serbyn_RMP2019,li_fisher_quantumzeno,skinner_nahum_prx2019,Li_Chen_Fisher}, as well as practical issues in quantum information~\cite{Nielsen_Chuang_2010,dynamics_qi_nat.rev.phy2019} and classical simulation~\cite{vidal_mps_prl2003,verstraete_cirac_mps_prl2004,white_feiguin_tebd_prl2004,daley_vidal_tebd2004,Eisert_Plenio_RMP2010}. Understanding its universal features helps characterize out-of-equilibrium phases and also constrains how efficiently quantum systems can be simulated or controlled.

A useful framework for understanding entanglement dynamics in many quantum systems with short-ranged interactions is the \textit{entanglement membrane} \cite{Nahum_Haah_entanglementgrowth_2016,Jonay2018,Zhou2019, Zhou2020,sierant_turkeshi_membrane2023}. Inspired by the Ryu-Takayanagi conjecture in AdS/CFT~\cite{Ryu_Takayanagi_2006}, this picture relates the entanglement entropy of a subsystem to the free energy of a membrane that stretches across spacetime, separating the subsystem from its complement, with the full system in a pure state. Rigorous mappings to classical statistical mechanics have validated this picture in specific quantum circuit models, such as Haar-random unitary circuits~\cite{Nahum_Haah_entanglementgrowth_2016,Zhou2019,Zhou2020,Random_Quantum_Circuits_review2023}, multi-unitary circuits~\cite{Foligno_Kos_Bertini_prl2024, Rampp_Claeys_PRR2024,Claeys_lamacraft_2024,liu_ho_prr2025}, and holographic models~\cite{vasseur_ludwig_PRB2019,mezei_prd2018,Mezei_jhep2020,jiang2024}, and have also been extended to broader classes of dynamics, including chaotic Hamiltonians~\cite{Jonay2018,Zhou2020} and non-unitary circuits~\cite{ippoliti_rakovszky_khemani_prx2022,Lovas_vijay_prxq2024} with measurements in the volume-law phase~\cite{jian_ludwig_prb2020,bao_choi_altman_prb2020,Li_Vijay_Fisher_DPRE2023}. It was recently shown that Clifford circuits realize a zero-temperature entanglement membrane~\cite{Sommers_Huse_2024}.

The membrane picture of entanglement growth suggests that higher-dimensional systems can exhibit qualitatively different behavior that does not occur in lower dimensions. This intuition arises from the classical statistical mechanics of membranes subjected to the periodic potential due to a lattice, as well as quenched disorder. In low dimensions, even weak disorder is a relevant perturbation that destabilizes lattice pinning and leads to disorder-dominated, rough membranes~\cite{Huse_Henley_1985,Huse_Henley_Fisher_respond_1985}. This explains why entanglement growth in (1+1)d random quantum circuits belongs to the Kardar-Parisi-Zhang (KPZ) universality class~\cite{Nahum_Haah_entanglementgrowth_2016,Zhou2019,Li_Vijay_Fisher_DPRE2023}. 
In contrast, higher-dimensional membranes can remain pinned to the lattice despite weak disorder. In particular, (3+1)d systems can exhibit a roughening transition, from smooth, lattice-pinned membranes to a rough, disorder-pinned phase~\cite{Halpin-Healy_FRG_1990,Emig_Nattermann_1998_FRGprl,Emig_Nattermann_1998_FRGlong,Noh_Rieger_2002_numerics}, with a rigorous mathematical proof of the smooth phase established in~\cite{Peled_mathproof2023} (see also~\cite{Bovier-Külske_3d,Bovier-Külske_2d,peled_2023_math_randomsurface}).  Motivated by this, we explore the 
entanglement growth in (3+1)d quantum circuits that exhibit an analogous randomness-induced roughening transition.

To probe this transition, we examine how entanglement entropy behaves across different spatial bipartitions after evolving the system under a  Clifford circuit with disorder. We introduce a \textit{hyperdiamond circuit} architecture that facilitates this analysis. Since the path of the membrane is not directly observable and is only accessible through its boundary, we diagnose roughening by examining how the membrane’s free energy---reflected in the entanglement entropy---responds to a tilt away from the lattice-pinned (smooth phase) orientation.  We develop a scaling theory that captures the crossover induced by this tilt, analogous to the rounding of a quantum critical point into a ``critical fan'' at finite temperature~\cite{Sachdev_2011} or finite evolution time~\cite{Feldmeier2021}. This framework allows us to identify distinct scaling regimes and extract critical behavior associated with the roughening transition.

\emph{Method---}
\begin{figure*}[tb]
    \centering
    \includegraphics[width=\textwidth]{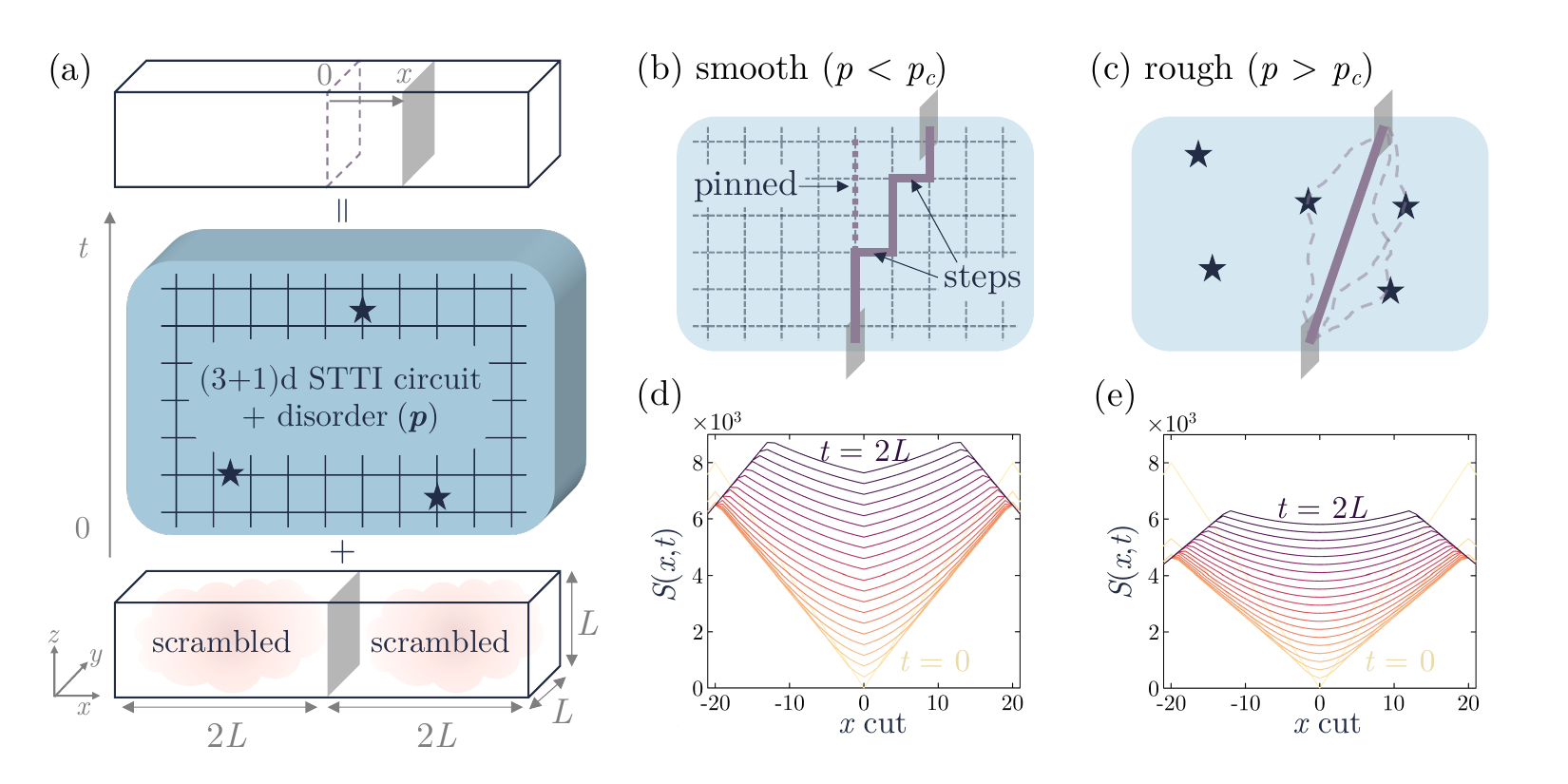}
    \caption{
    (a) Illustration of the setup and protocol. The initial state consists of two independent subsystems ($x<0$ and $x>0$), each with $2L^3$ qubits prepared in a near-maximally entangled pure state via a random unitary Clifford circuit. 
    The entanglement membrane is initially anchored at the central cut ($x=0$) across which there is no entanglement at $t=0$. Time evolution is then performed using a 3+1-dimensional STTI Clifford circuit (illustrated as a lattice) perturbed by disorder (indicated by stars) with rate $p$. After time $t$, we compute the bipartite entanglement entropy $S(x,t)$ across cuts perpendicular to $\hat{x}$ and average this over disorder realizations. These cuts also serve as anchoring surfaces for the entanglement membrane at $(x,t)$.
    (b) In the smooth phase ($p<p_c$), the membrane remains pinned to the lattice.  Moving the cut to $x \neq 0$ forces the membrane to take $|x|$ discrete steps, each contributing to the energy, resulting in a cusp in $S(x,t)$ at $x=0$, as in (d).
    (c) In the rough phase ($p>p_c$), the membrane becomes pinned by disorder, so it is rough on large scales. The sample-averaged entanglement $S(x,t)$ smooths out near $x=0$, reflecting the minimal volume connecting the anchoring surfaces.
    (d) Results for $p=0.05<p_c$ show the characteristic cusp at $x=0$.
    (e) Numerical results in the rough phase ($p=0.15>p_c$) show a smooth quadratic profile near $x=0$ without a cusp.
    All numerical results in this figure are for system size $L = 20$, with time evolution shown at even times up to $t = 2L$. 
    }
    \label{fig:schematics}
\end{figure*}
As illustrated in Fig.~\ref{fig:schematics}(a), we initialize the system in a stabilizer state composed of two independently scrambled halves. Each half consists of $2L^3$ qubits, prepared in a near-maximally entangled pure state~\footnote{We prepare a near-maximally entangled state by acting with a global random unitary Clifford circuit on the product state $|0\rangle^{\otimes 2L^3}$. The resulting state is a Page-like state.}.
The two subsystems are then joined along the $x = 0$ plane to form a full system of size $4L \times L \times L$, with $\hat{x}$ as the long axis. Because the two halves are prepared independently, the initial bipartite entanglement across the central cut vanishes, thus pinning one end of the entanglement membrane to $(x=0,t=0)$.

We then evolve the state under a Clifford circuit that satisfies open boundary conditions along the $\hat{x}$ direction while the boundary conditions are periodic along the other two directions. After time $t$, we examine the bipartite entanglement entropy $S(x,t)$ across cuts at constant $x\in\mathbb{Z}$. The entanglement is governed by the minimal energy of a $(2+1)$d membrane spanning the circuit between these two-dimensional ``anchoring'' surfaces (cuts): one at $(x=0, t=0)$ and the other at $(x, t)$.  Since the membrane in a Clifford circuit corresponds to an effective zero-temperature limit \cite{Sommers_Huse_2024}, it is an effective energy rather than a free energy that is minimized.  The scaling of $S(x,t)$ with $x$ and $t$ then corresponds to the minimal energy membrane configuration under these boundary conditions.

The circuit we study is built from a space-time-translationally-invariant (STTI)~\cite{Sommers_Huse_Gullans_2023,Sommers_Huse_2024} structure forming a regular lattice. Disorder is additionally introduced by locally changing the gates with probability $p$, thereby breaking the space-time lattice symmetry in a controlled, stochastic way. Without disorder, $p=0$, the circuit forms a STTI lattice, which selects preferred \textit{lattice-pinned} membrane orientations that minimize the entanglement entropy density.  We orient the cuts so that the entanglement membrane at $x=0$ is one of these lattice-pinned orientations [see Fig.~\ref{fig:schematics}(b), purple dotted line].  If the cut at $t>0$ is moved to $x\neq0$, the membrane must take $|x|$ discrete steps to reach that cut.  In the {\it smooth phase}, $p<p_c$, each step contributes a free energy cost proportional to its area $L^2$, so for $|x|\ll t$
\begin{align}
    \label{eqn:smooth}
    S(x,t)-S(0,t)\sim|x|L^2~.
\end{align}
This produces a characteristic cusp at $x=0$ in the entanglement profile $S(x,t)$ as in Fig.~\ref{fig:schematics}(d). Notably, this cusp persists for small nonzero disorder rates $p<p_c$, where the membrane remains smooth and pinned to the underlying lattice.

For stronger disorder $p > p_c$, the membrane is no longer pinned by the lattice but instead becomes \textit{pinned by disorder} on large scales [Fig.~\ref{fig:schematics}(c)]. Hence, the membrane is \textit{rough}. In this regime, the membrane asymptotically ignores the underlying lattice, and the leading contribution to the disorder-averaged entanglement entropy comes from the minimal volume connecting the initial and final anchoring surfaces. This gives
\begin{align}
    \label{eqn:rough}
    S(x,t)-S(0,t)\sim (\sqrt{t^2+x^2}-t)L^2\sim x^2L^2/t~,
\end{align}
so the entanglement profile becomes smooth near $x = 0$, scaling quadratically with $x$ [see Fig.~\ref{fig:schematics}(e)].

\emph{The hyperdiamond circuit---}
As discussed earlier, we choose the circuit geometry such that $\hat{x}$ is normal to one of the membrane's pinned orientations. We also require the circuit to have a finite butterfly velocity $v_B < \infty$ to avoid pathological cases where entanglement fails to grow~\footnote{This is why we do not use the simple hypercubic lattice with time running along $\hat{t} = (1,1,1,1)$; such geometries typically lead to infinite butterfly velocity, which can artificially suppress entanglement growth~\cite{Sommers_Huse_2024}. We avoid this fine-tuned limit to focus on the roughening physics.}. In addition, we prefer the circuit geometry to be isotropic in both space and time. To satisfy all these conditions, we design a \textit{hyperdiamond} circuit architecture.

Consider qubits placed on a three-dimensional Cartesian grid labeled by $\textbf{r} \in \mathbb{Z}^3$, subject to a choice of boundary conditions. For each lattice site $\textbf{r}$, we define a set of target sites, $\mathcal{S}_{T}(\textbf{r}) = \{\textbf{r}+\textbf{e}_i|i\in\{1,2,3\}\}$, where $\textbf{e}_i$ are the unit vectors along the three coordinate directions. If $\textbf{r} + \textbf{e}_i$ lies outside the system under open boundary conditions, the corresponding direction is omitted from the set. For periodic boundaries, the location is taken modulo the system size in that direction, $L$.

Our circuit consists of one- and two-qubit gates, which are independently and randomly selected for each realization.  Two-qubit gates are applied at integer times $t \in \mathbb{Z}$. The one-qubit gate $G_1(t,\textbf{r})$ acts between times $t$ and $(t+1)$ on qubit $\textbf{r}$, and is set to be a Hadamard gate with probability $p$ or the identity gate with probability $1-p$. Similarly, the two-qubit gate $G_2(t,\textbf{r}_C,\textbf{r}_T)$ acts at time $t$, with $\textbf{r}_C$ as the control and $\textbf{r}_T$ as the target. With probability $1-p$, $G_2$ applies a $\text{CNOT}(\textbf{r}_C,\textbf{r}_T)$ gate, and with probability $p$, the ``gate'' $G_2$ performs a projective measurement of the Pauli operator $\hat{Z}(\textbf{r}_C)\hat{X}(\textbf{r}_T)$ for that circuit realization.

Once the gates are predetermined for a given realization, the total circuit up to time $T$ is 
\begin{align}
    U(T,p) \equiv 
    \mathcal{T}\Big[\prod_{t=1}^T
    \Big(
    \prod_{\textbf{r}}G_1(t,\textbf{r})
    \prod_{\substack{\textbf{r}\equiv t(\text{mod}2) \\ \textbf{r}' \in \mathcal{S}_{T}(\textbf{r})}} 
    G_2(t,\textbf{r},\textbf{r}')
    \Big) \Big]~,
\end{align}
where we alternate the two-qubit gates on odd and even control sites, so that one full period consists of two time steps, $T_p = 2$. The parity of a site $\textbf{r} = (x, y, z)$ is defined by the parity of $(x + y + z)$, and $\mathcal{T}[\cdot]$ denotes time-ordering of the gates.

In the disorder-free limit $p=0$, the circuit reduces to a fully deterministic unitary evolution: 
\begin{align}
    U(T_p,0) =
    \prod_{\substack{\text{even}~\textbf{r} \\ \textbf{r}' \in \mathcal{S}_{T}(\textbf{r})}} 
    \text{CNOT}(\textbf{r},\textbf{r}')
    \prod_{\substack{\text{odd}~\textbf{r} \\ \textbf{r}' \in \mathcal{S}_{T}(\textbf{r})}} 
    \text{CNOT}(\textbf{r},\textbf{r}')~.
\end{align}
This defines an STTI multi-unitary circuit that can also be represented as being on a $3+1=4$-dimensional hyperdiamond lattice, with the full point group symmetry of that 4d lattice.  In the ZX-calculus language, this hyperdiamond circuit consists of $Z$-spiders with Hadamard gates on all edges (see Supplementary Material for details~\cite{supp_mat}), generalizing the well-known honeycomb lattice in 2d and the diamond lattice in 3d. At $p>0$, in this hyperdiamond/ZX-calculus representation of our circuit, a fraction $p$ of the one-qubit Hadamards are replaced with identity gates isotropically in space and time.  In the hypercubic representation, this replaces CNOT gates by $\hat{Z}\hat{X}$ measurements on the spacelike legs.

\emph{Scaling Argument---}
\begin{figure}[tb]
    \centering
    \includegraphics[width=0.4\textwidth]{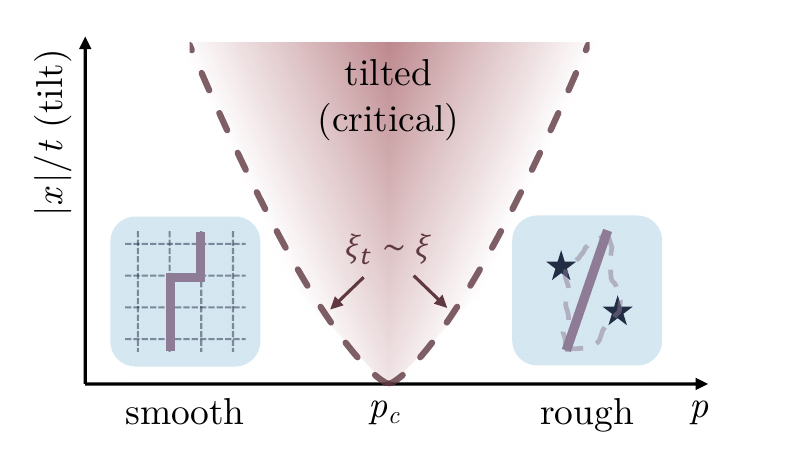}
    \caption{Scaling regimes of a 3d interface with tilt $|x|/t$ and disorder strength $p$. $\xi \sim |p-p_c|^{-\nu}$ is the correlation length in an infinite system. $\xi_t = t/|x|$ is the length scale beyond which we ``notice'' the tilt. The crossovers between the tilted critical regime and the rough or smooth phases occur at $\xi_t \sim \xi$.  The roughening phase transition at $p=p_c$ only occurs at zero tilt.
    }
    \label{fig:scaling}
\end{figure}
Consider evolving the system under this circuit for a time proportional to the system size ($t \sim L$).
We present the scaling arguments with $t = 2L$, to match the numerics that we present below.  While the precise form of the scaling function may depend on the value of $t/L$, the scaling exponents and the overall structure of the scaling relations do not depend on this ratio.

Without disorder, the entanglement membrane from the center cut ($x=0$) spans a volume of $t\times (t/2)\times (t/2)$, perfectly pinned by the lattice. Including disorder, membrane displacements along the $\hat{x}$ direction are suppressed by the lattice pinning and scale as $O(t^0)$ in the smooth phase $p<p_c$, while they grow as $O(t^{\zeta_r})$ with the roughening exponent $\zeta_r>0$ in the rough phase $p>p_c$.
However, in these quantum circuit calculations, we do not have direct access to the membrane configuration within the bulk of the circuit \footnote{One could try using an ancilla as a probe to detect some aspects of the membrane geometry as mentioned in \cite{Sommers_Huse_2024}.}. The quantity that we explore instead is the entanglement profile $S(x,t)$, in particular for $t=2L$. 
For nonzero $x$ this imposes a boundary condition that forces the membrane to deviate from the lattice-pinned direction.  We develop a scaling theory for the entanglement entropy under such an imposed tilt of the membrane in the vicinity of the roughening transition: $p$ near $p_c$.

We particularly focus on $\delta S(x,t,p)\equiv S(x,t) - t^3 s_a(p) - t^2 s_b(p)$ where $s_a(p)$ and $s_b(p)$ are the analytic backgrounds at $x=0$ originating from the membrane volume $\sim t^3$, and the pinned boundary at the upper and the lower edges $\sim t^2$, respectively. We write the scaling ansatz
\begin{align}
    \delta S(x,t,p) \approx t^{\theta_c} F\left(x,(p-p_c) t^{1/\nu}\right),
    \label{eqn:scaling}
\end{align}
where $F(x,y)$ is the scaling function (which also depends on $t/L$), $\theta_c$ is the violation of hyperscaling exponent at the transition~\cite{Emig_Nattermann_1998_FRGprl}, and $\nu$ is the correlation length exponent: the correlation length is $\xi\sim|p-p_c|^{-\nu}$. Note that $|y|\sim(t/\xi)^{1/\nu}$.
The first argument of the scaling function is independent of $t$, as at the roughening transition the membrane is logarithmically rough \cite{Emig_Nattermann_1998_FRGprl,Emig_Nattermann_1998_FRGlong,Noh_Rieger_2002_numerics}, with no time-dependent characteristic displacement normal to its mean position.  The functional renormalization group (FRG) calculation predicts $\theta_c=3/2$ for our case of $d=3$ at the leading order in $\epsilon=4-d$~\cite{Emig_Nattermann_1998_FRGprl}.

At the critical point $p=p_c$ (or $y=0$), for small tilts $t\gg|x|\gg1$, we expect
\begin{align}
\label{eqn:scaling_critical}
    F(x,0) - F(0,0) \approx B_c|x|^{3 - \theta_c}~,
\end{align}
with $B_c>0$.
The exponent is determined to connect smoothly with the limit of $x\sim t$, where $\delta S$ should be proportional to the membrane volume $\sim t^3$. We also expect the exponent to be in the interval $3-\theta_c\in(1,2)$, intermediate between smooth [Eqn.~(\ref{eqn:smooth})] and rough [Eqn.~(\ref{eqn:rough})] phases.

We next focus on the scaling function in both phases for zero and small tilt $|x|$, where we will quantify the crossover later. Note that $x$ is always an integer, and the lattice constant provides the UV cutoff.
In the smooth phase ($y \ll -1$), 
\begin{align}
    \label{eqn:scaling_smooth}
    F(x,y\ll-1) \approx A_s |y|^{(3 - \theta_c)\nu} + B_s |x| |y|^{(2 - \theta_c)\nu}~,
\end{align}
with $A_s > 0$ and $B_s > 0$. The first term matches with the expected scaling $\delta S(x=0,p <p_c) \sim t^3$ when $t \gg \xi$. 
The second term accounts for the $x$-dependence due to steps in the membrane, consistent with Eqn.~(\ref{eqn:smooth}).
The scaling function in the rough phase ($y \gg 1$) behaves as
\begin{align}
    \label{eqn:scaling_rough}
    F(x,y\gg1) \approx A_r y^{(\theta_r - \theta_c)\nu} + B_r x^2 y^{(1 - \theta_c)\nu}~,
\end{align}
with $A_r < 0$ and $B_r > 0$. The first term arises from $\delta S (x=0,p>p_c) \sim t^{\theta_r}$, with free-energy roughening exponent $\theta_r$, as the system at large scale $t\gg\xi$ gets rougher to lower the entanglement entropy by finding the lowest-entropy positions, suggesting negative $A_r$. The second term matches with Eqn.~(\ref{eqn:rough}).

The membrane is pinned by the cuts at both $(0,0)$ and $(x,t)$, so the characteristic temporal scale for detecting a tilt by one lattice spacing away from the smooth ($x=0$) orientation is $\sim t/|x|$, which we define as $\xi_t$. The scaling form for ``small enough'' $|x|$ corresponds to $\xi_t \gg \xi$, i.e., when the tilt is too small to be resolved within one correlation length. The crossover to the tilted regime occurs when $\xi_t \sim \xi$, or equivalently $x \sim y^\nu$.
For $p$ near $p_c$, Eqns.~(\ref{eqn:scaling_smooth}) and (\ref{eqn:scaling_rough}) apply in the weakly tilted regime $\xi_t\gg\xi$, while Eqn.~(\ref{eqn:scaling_critical}) applies in the more strongly tilted critical regime $\xi\gg\xi_t$, with these regimes matching smoothly at the crossover for any nonzero small tilt.
The resulting phase diagram, with a sharp roughening transition only at zero tilt, is shown in Fig.~\ref{fig:scaling}. We refer readers to the Supplementary Material for further details on the scaling relations~\cite{supp_mat}. 

\begin{figure*}[t]
    \centering
    \includegraphics[width=\textwidth]{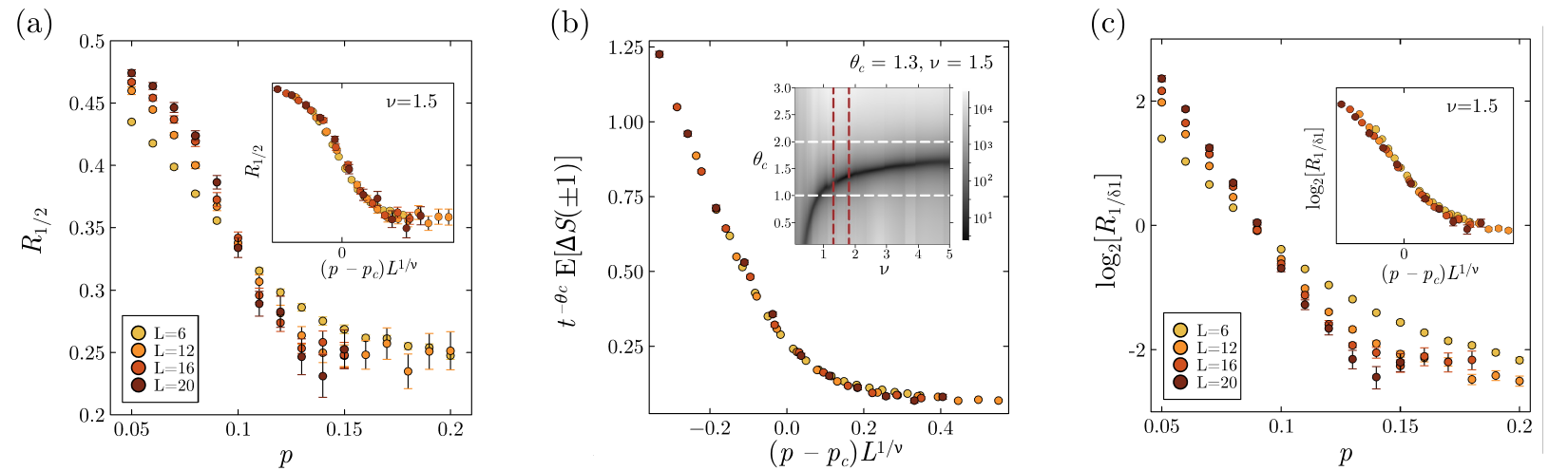}
    \caption{Numerical results from the hyperdiamond circuit at runtime $t = 2L$ and associated finite-size scaling analysis.
    (a) The ratio $R_{1/2} \equiv \mathbb{E}[\Delta S(\pm 1)] / \mathbb{E}[\Delta S(\pm 2)]$ shows a crossing near $p_c \approx 0.095$. The inset shows scaling collapse with $\nu = 1.5$.
    (b) Finite-size scaling of $\mathbb{E}[\Delta S(\pm 1)]$, using the previously estimated $\nu \approx 1.5$. The best collapse yields $\theta_c \approx 1.3$. The inset shows the objective function evaluating collapse quality across $\theta_c$ and $\nu$. The white dotted line marks the expected regime $\theta_c \in (1, 2)$, and the brown dotted lines indicate the bounds on $\nu$ inferred from panel (a).
    (c) The ratio $R_{1/\delta 1} \equiv \mathbb{E}[\Delta S(\pm1)] / \sigma[\Delta S(\pm1)]$ also exhibits a crossing near $p_c \approx 0.095$, with a scaling collapse shown as the inset.
    }
    \label{fig:numerics}
\end{figure*}

\emph{Numerical Results---} 
As in the previous section, we focus on circuit run times proportional to the system size, with $t = 2L$. 
We define the entanglement difference from the center as $\Delta S(x) \equiv S(x,t) - S(0,t)$. To estimate the roughening transition, we consider the dimensionless ratio $R_{1/2} \equiv \mathbb{E}[\Delta S(\pm1)] / \mathbb{E}[\Delta S(\pm2)]$, where $\mathbb{E}[\cdot]$ denotes the average over disorder realizations. In the thermodynamic limit, this ratio approaches $1/2$ in the smooth phase and $1/4$ in the rough phase. Fig.~\ref{fig:numerics}(a) shows $R_{1/2}$ as a function of $p$ for various system sizes. For larger sizes ($L \geq 16$), we average over a short time window near $t = L$ to reduce statistical noise in $\Delta S$. In the rough phase at larger $p$, sample-to-sample fluctuations in $\Delta S$ become comparable to the mean, introducing significant uncertainty in estimating the critical behavior. While this limits the precision of a full finite-size scaling analysis, the crossings of $R_{1/2}$ still reflect universal features of the scaling function—specifically, the relative values of $F(0,0)$, $F(1,0)$, and $F(2,0)$—and provide a robust estimate of the critical point. From these crossings, we estimate the critical rate to be $p_c \approx 0.095$, located between $p = 0.09$ and $p = 0.1$. 
The scaling collapse shown in the inset of Fig.~\ref{fig:numerics}(a) yields a correlation length exponent of $\nu = 1.5 \pm 0.3$, consistent with previous results from solid-on-solid models~\cite{Noh_Rieger_2002_numerics}. Our numerics use up to $7 \times 10^4$ disorder realizations for smaller systems ($L = 6$), especially in the rough phase to suppress statistical errors, and approximately $10^3$ samples for the largest system size $L = 20$.

Next, we extract the exponent $\theta_c$ by examining the free energy cost of a single step ($|x| = 1$), given by $\mathbb{E}[\Delta S(\pm1)]$. Based on the scaling form in Eqn.~(\ref{eqn:scaling}), we expect a data collapse when plotting $t^{-\theta_c} \mathbb{E}[\Delta S(\pm1)]$ vs. $|p - p_c| L^{1/\nu}$. To assess the quality of the collapse, we compute an objective function~\cite{scaling} shown as an inset in Fig.~\ref{fig:numerics}(b); darker regions indicate better collapse. Due to limitations from finite system sizes, the 68\% confidence interval on the critical exponent is not strictly bounded. Nevertheless, using the estimated bounds on $\nu$ inferred from $R_{1/2}$, we estimate $\theta_c \approx 1.3 \pm 0.2$.

We additionally consider the ratio between the mean and standard deviation, $R_{1/\delta 1} \equiv \mathbb{E}[\Delta S(\pm1)] / \sigma[\Delta S(\pm1)]$, shown in Fig.~\ref{fig:numerics}(c), which helps suppress statistical fluctuations in the rough phase. $\sigma[\cdot]$ denotes the standard deviation over samples. The ratio is dimensionless at the critical point, so we expect crossings across different system sizes. In the smooth phase, $R_{1/\delta 1} \sim O(t^{1/2})$, while in the rough phase $R_{1/\delta 1} \sim O(1)$. This scaling arises because the difference between entanglement membranes for $S(x=1)$ and $S(x=0)$ corresponds to a volume $O(t^3)$ in the smooth phase, but only an anchored area $O(t^2)$ in the rough phase. Therefore, $\sigma[\Delta S(\pm1)]$ scales as $O(t^{3/2})$ in the smooth phase and $O(t)$ in the rough phase.  The crossing again occurs between $p = 0.09$ and $p = 0.1$, consistent with the $R_{1/2}$ results. However, the scaling collapse is of poorer quality as compared to that for $R_{1/2}$.

\emph{Discussion---}
In this work, we demonstrated that the periodic lattice structure of a quantum circuit can significantly influence entanglement growth, and that this effect can persist in the presence of disorder in high dimensions. Specifically, we characterized the roughening transition in a $(3+1)$d Clifford circuit model by tuning the degree of randomness in the \textit{hyperdiamond} circuit, which we designed to efficiently access the relevant physics. By preparing an initial state that anchors the entanglement membrane at the center, we probed entanglement across various spatial cuts. This corresponds to tilting the membrane away from the lattice-pinned orientation and studying the resulting free energy cost associated with the tilt.
We additionally developed a scaling theory incorporating such tilt effects, which rounds the sharp critical behavior of the transition and leads to a crossover regime. From scaling collapse, we extracted the correlation length exponent $\nu\approx1.5$, consistent with previous numerical studies, and estimated the critical exponent $\theta_c\approx1.3$, which had not been determined previously. While a value for $\theta_c$ has been predicted theoretically using a perturbative FRG approach~\cite{Emig_Nattermann_1998_FRGlong}, our work provides what appears to be the first direct numerical estimate of this exponent.

High-dimensional and translationally invariant circuit architectures are both timely and important. Such geometries have attracted attention for enabling robust quantum memories and fault-tolerant computation~\cite{dennis2002,Breuckmann2017}, e.g., the 4d toric code is thermally stable~\cite{horodecki_4dtoric2010}. They also support better quantum error-correcting codes to overcome strict bounds on code rate in lower dimensions~\cite{bpt_bound_prl2010,Flammia_quantum2017,Breuckmann_prxq2021,tremblay_prl2022,Bravyi_nature2024}. On the experimental side, reconfigurable platforms now make it possible to simulate these extended geometries effectively~\cite{Bluvstein_nature2022,Bluvstein_nature2024,xu_nature2024}. Translational invariance is similarly motivated: native gate operations can be applied in parallel, naturally giving rise to STTI circuits. These setups not only reflect realistic experimental constraints but also offer scalable routes to studying quantum dynamics and computation. Our work addresses this regime, providing a concrete framework to understand how entanglement grows and transitions in such high-dimensional and structured systems, where both theoretical insight and experimental timeliness come together.

\section{Acknowledgments}  
We thank Tyler Cochran, Sarang Gopalakrishnan, Michael Knap, Su-un Lee, Yaodong Li, Bendeguz Offertaler, Ron Peled, Rhine Samajdar, and Yifan Zhang for valuable discussions.
  This work was supported in part by NSF QLCI grant OMA-2120757.  Numerical work was completed using computational resources managed and supported by Princeton Research Computing, a consortium of groups including the Princeton Institute for Computational Science and Engineering (PICSciE) and the Office of Information Technology's High Performance Computing Center and Visualization Laboratory at Princeton University. The open-source \texttt{QuantumClifford.jl} package was used to simulate Clifford circuits~\cite{QuantumClifford}.

\bibliography{main}

\begin{thebibliography}{89}%
\makeatletter
\providecommand \@ifxundefined [1]{%
 \@ifx{#1\undefined}
}%
\providecommand \@ifnum [1]{%
 \ifnum #1\expandafter \@firstoftwo
 \else \expandafter \@secondoftwo
 \fi
}%
\providecommand \@ifx [1]{%
 \ifx #1\expandafter \@firstoftwo
 \else \expandafter \@secondoftwo
 \fi
}%
\providecommand \natexlab [1]{#1}%
\providecommand \enquote  [1]{``#1''}%
\providecommand \bibnamefont  [1]{#1}%
\providecommand \bibfnamefont [1]{#1}%
\providecommand \citenamefont [1]{#1}%
\providecommand \href@noop [0]{\@secondoftwo}%
\providecommand \href [0]{\begingroup \@sanitize@url \@href}%
\providecommand \@href[1]{\@@startlink{#1}\@@href}%
\providecommand \@@href[1]{\endgroup#1\@@endlink}%
\providecommand \@sanitize@url [0]{\catcode `\\12\catcode `\$12\catcode `\&12\catcode `\#12\catcode `\^12\catcode `\_12\catcode `\%12\relax}%
\providecommand \@@startlink[1]{}%
\providecommand \@@endlink[0]{}%
\providecommand \url  [0]{\begingroup\@sanitize@url \@url }%
\providecommand \@url [1]{\endgroup\@href {#1}{\urlprefix }}%
\providecommand \urlprefix  [0]{URL }%
\providecommand \Eprint [0]{\href }%
\providecommand \doibase [0]{https://doi.org/}%
\providecommand \selectlanguage [0]{\@gobble}%
\providecommand \bibinfo  [0]{\@secondoftwo}%
\providecommand \bibfield  [0]{\@secondoftwo}%
\providecommand \translation [1]{[#1]}%
\providecommand \BibitemOpen [0]{}%
\providecommand \bibitemStop [0]{}%
\providecommand \bibitemNoStop [0]{.\EOS\space}%
\providecommand \EOS [0]{\spacefactor3000\relax}%
\providecommand \BibitemShut  [1]{\csname bibitem#1\endcsname}%
\let\auto@bib@innerbib\@empty
\bibitem [{\citenamefont {Amico}\ \emph {et~al.}(2008)\citenamefont {Amico}, \citenamefont {Fazio}, \citenamefont {Osterloh},\ and\ \citenamefont {Vedral}}]{Amico_Vedral_RMP2008}%
  \BibitemOpen
  \bibfield  {author} {\bibinfo {author} {\bibfnamefont {L.}~\bibnamefont {Amico}}, \bibinfo {author} {\bibfnamefont {R.}~\bibnamefont {Fazio}}, \bibinfo {author} {\bibfnamefont {A.}~\bibnamefont {Osterloh}},\ and\ \bibinfo {author} {\bibfnamefont {V.}~\bibnamefont {Vedral}},\ }\bibfield  {title} {\bibinfo {title} {Entanglement in many-body systems},\ }\href {https://doi.org/10.1103/RevModPhys.80.517} {\bibfield  {journal} {\bibinfo  {journal} {Rev. Mod. Phys.}\ }\textbf {\bibinfo {volume} {80}},\ \bibinfo {pages} {517} (\bibinfo {year} {2008})}\BibitemShut {NoStop}%
\bibitem [{\citenamefont {Page}(1993)}]{page_prl1993}%
  \BibitemOpen
  \bibfield  {author} {\bibinfo {author} {\bibfnamefont {D.~N.}\ \bibnamefont {Page}},\ }\bibfield  {title} {\bibinfo {title} {Average entropy of a subsystem},\ }\href {https://doi.org/10.1103/PhysRevLett.71.1291} {\bibfield  {journal} {\bibinfo  {journal} {Phys. Rev. Lett.}\ }\textbf {\bibinfo {volume} {71}},\ \bibinfo {pages} {1291} (\bibinfo {year} {1993})}\BibitemShut {NoStop}%
\bibitem [{\citenamefont {Kim}\ and\ \citenamefont {Huse}(2013)}]{Kim_Huse_2013}%
  \BibitemOpen
  \bibfield  {author} {\bibinfo {author} {\bibfnamefont {H.}~\bibnamefont {Kim}}\ and\ \bibinfo {author} {\bibfnamefont {D.~A.}\ \bibnamefont {Huse}},\ }\bibfield  {title} {\bibinfo {title} {Ballistic spreading of entanglement in a diffusive nonintegrable system},\ }\href {https://doi.org/10.1103/PhysRevLett.111.127205} {\bibfield  {journal} {\bibinfo  {journal} {Phys. Rev. Lett.}\ }\textbf {\bibinfo {volume} {111}},\ \bibinfo {pages} {127205} (\bibinfo {year} {2013})}\BibitemShut {NoStop}%
\bibitem [{\citenamefont {D'Alessio}\ \emph {et~al.}(2016)\citenamefont {D'Alessio}, \citenamefont {Kafri}, \citenamefont {Polkovnikov},\ and\ \citenamefont {and}}]{d'alessio_rigol_review2016}%
  \BibitemOpen
  \bibfield  {author} {\bibinfo {author} {\bibfnamefont {L.}~\bibnamefont {D'Alessio}}, \bibinfo {author} {\bibfnamefont {Y.}~\bibnamefont {Kafri}}, \bibinfo {author} {\bibfnamefont {A.}~\bibnamefont {Polkovnikov}},\ and\ \bibinfo {author} {\bibfnamefont {M.~R.}\ \bibnamefont {and}},\ }\bibfield  {title} {\bibinfo {title} {From quantum chaos and eigenstate thermalization to statistical mechanics and thermodynamics},\ }\href {https://doi.org/10.1080/00018732.2016.1198134} {\bibfield  {journal} {\bibinfo  {journal} {Advances in Physics}\ }\textbf {\bibinfo {volume} {65}},\ \bibinfo {pages} {239} (\bibinfo {year} {2016})},\ \Eprint {https://arxiv.org/abs/https://doi.org/10.1080/00018732.2016.1198134} {https://doi.org/10.1080/00018732.2016.1198134} \BibitemShut {NoStop}%
\bibitem [{\citenamefont {Kaufman}\ \emph {et~al.}(2016)\citenamefont {Kaufman}, \citenamefont {Tai}, \citenamefont {Lukin}, \citenamefont {Rispoli}, \citenamefont {Schittko}, \citenamefont {Preiss},\ and\ \citenamefont {Greiner}}]{kaufman_science2016}%
  \BibitemOpen
  \bibfield  {author} {\bibinfo {author} {\bibfnamefont {A.~M.}\ \bibnamefont {Kaufman}}, \bibinfo {author} {\bibfnamefont {M.~E.}\ \bibnamefont {Tai}}, \bibinfo {author} {\bibfnamefont {A.}~\bibnamefont {Lukin}}, \bibinfo {author} {\bibfnamefont {M.}~\bibnamefont {Rispoli}}, \bibinfo {author} {\bibfnamefont {R.}~\bibnamefont {Schittko}}, \bibinfo {author} {\bibfnamefont {P.~M.}\ \bibnamefont {Preiss}},\ and\ \bibinfo {author} {\bibfnamefont {M.}~\bibnamefont {Greiner}},\ }\bibfield  {title} {\bibinfo {title} {Quantum thermalization through entanglement in an isolated many-body system},\ }\href {https://doi.org/10.1126/science.aaf6725} {\bibfield  {journal} {\bibinfo  {journal} {Science}\ }\textbf {\bibinfo {volume} {353}},\ \bibinfo {pages} {794} (\bibinfo {year} {2016})},\ \Eprint {https://arxiv.org/abs/https://www.science.org/doi/pdf/10.1126/science.aaf6725} {https://www.science.org/doi/pdf/10.1126/science.aaf6725} \BibitemShut {NoStop}%
\bibitem [{\citenamefont {Bianchi}\ \emph {et~al.}(2022)\citenamefont {Bianchi}, \citenamefont {Hackl}, \citenamefont {Kieburg}, \citenamefont {Rigol},\ and\ \citenamefont {Vidmar}}]{bianchi_vidmar_prxq2022}%
  \BibitemOpen
  \bibfield  {author} {\bibinfo {author} {\bibfnamefont {E.}~\bibnamefont {Bianchi}}, \bibinfo {author} {\bibfnamefont {L.}~\bibnamefont {Hackl}}, \bibinfo {author} {\bibfnamefont {M.}~\bibnamefont {Kieburg}}, \bibinfo {author} {\bibfnamefont {M.}~\bibnamefont {Rigol}},\ and\ \bibinfo {author} {\bibfnamefont {L.}~\bibnamefont {Vidmar}},\ }\bibfield  {title} {\bibinfo {title} {Volume-law entanglement entropy of typical pure quantum states},\ }\href {https://doi.org/10.1103/PRXQuantum.3.030201} {\bibfield  {journal} {\bibinfo  {journal} {PRX Quantum}\ }\textbf {\bibinfo {volume} {3}},\ \bibinfo {pages} {030201} (\bibinfo {year} {2022})}\BibitemShut {NoStop}%
\bibitem [{\citenamefont {Nandkishore}\ and\ \citenamefont {Huse}(2015)}]{nandkishore_huse_mbl}%
  \BibitemOpen
  \bibfield  {author} {\bibinfo {author} {\bibfnamefont {R.}~\bibnamefont {Nandkishore}}\ and\ \bibinfo {author} {\bibfnamefont {D.~A.}\ \bibnamefont {Huse}},\ }\bibfield  {title} {\bibinfo {title} {Many-body localization and thermalization in quantum statistical mechanics},\ }\href {https://doi.org/10.1146/annurev-conmatphys-031214-014726} {\bibfield  {journal} {\bibinfo  {journal} {Annual Review of Condensed Matter Physics}\ }\textbf {\bibinfo {volume} {6}},\ \bibinfo {pages} {15} (\bibinfo {year} {2015})}\BibitemShut {NoStop}%
\bibitem [{\citenamefont {Gopalakrishnan}\ and\ \citenamefont {Huse}(2019)}]{gopalakrishnan_huse_prb2019}%
  \BibitemOpen
  \bibfield  {author} {\bibinfo {author} {\bibfnamefont {S.}~\bibnamefont {Gopalakrishnan}}\ and\ \bibinfo {author} {\bibfnamefont {D.~A.}\ \bibnamefont {Huse}},\ }\bibfield  {title} {\bibinfo {title} {Instability of many-body localized systems as a phase transition in a nonstandard thermodynamic limit},\ }\href {https://doi.org/10.1103/PhysRevB.99.134305} {\bibfield  {journal} {\bibinfo  {journal} {Phys. Rev. B}\ }\textbf {\bibinfo {volume} {99}},\ \bibinfo {pages} {134305} (\bibinfo {year} {2019})}\BibitemShut {NoStop}%
\bibitem [{\citenamefont {Alet}\ and\ \citenamefont {Laflorencie}(2018)}]{alet_laflorencie_mbl_review}%
  \BibitemOpen
  \bibfield  {author} {\bibinfo {author} {\bibfnamefont {F.}~\bibnamefont {Alet}}\ and\ \bibinfo {author} {\bibfnamefont {N.}~\bibnamefont {Laflorencie}},\ }\bibfield  {title} {\bibinfo {title} {Many-body localization: An introduction and selected topics},\ }\href {https://doi.org/https://doi.org/10.1016/j.crhy.2018.03.003} {\bibfield  {journal} {\bibinfo  {journal} {Comptes Rendus Physique}\ }\textbf {\bibinfo {volume} {19}},\ \bibinfo {pages} {498} (\bibinfo {year} {2018})}\BibitemShut {NoStop}%
\bibitem [{\citenamefont {Abanin}\ \emph {et~al.}(2019)\citenamefont {Abanin}, \citenamefont {Altman}, \citenamefont {Bloch},\ and\ \citenamefont {Serbyn}}]{Abanin_Serbyn_RMP2019}%
  \BibitemOpen
  \bibfield  {author} {\bibinfo {author} {\bibfnamefont {D.~A.}\ \bibnamefont {Abanin}}, \bibinfo {author} {\bibfnamefont {E.}~\bibnamefont {Altman}}, \bibinfo {author} {\bibfnamefont {I.}~\bibnamefont {Bloch}},\ and\ \bibinfo {author} {\bibfnamefont {M.}~\bibnamefont {Serbyn}},\ }\bibfield  {title} {\bibinfo {title} {Colloquium: Many-body localization, thermalization, and entanglement},\ }\href {https://doi.org/10.1103/RevModPhys.91.021001} {\bibfield  {journal} {\bibinfo  {journal} {Rev. Mod. Phys.}\ }\textbf {\bibinfo {volume} {91}},\ \bibinfo {pages} {021001} (\bibinfo {year} {2019})}\BibitemShut {NoStop}%
\bibitem [{\citenamefont {Li}\ \emph {et~al.}(2018)\citenamefont {Li}, \citenamefont {Chen},\ and\ \citenamefont {Fisher}}]{li_fisher_quantumzeno}%
  \BibitemOpen
  \bibfield  {author} {\bibinfo {author} {\bibfnamefont {Y.}~\bibnamefont {Li}}, \bibinfo {author} {\bibfnamefont {X.}~\bibnamefont {Chen}},\ and\ \bibinfo {author} {\bibfnamefont {M.~P.~A.}\ \bibnamefont {Fisher}},\ }\bibfield  {title} {\bibinfo {title} {Quantum zeno effect and the many-body entanglement transition},\ }\href {https://doi.org/10.1103/PhysRevB.98.205136} {\bibfield  {journal} {\bibinfo  {journal} {Phys. Rev. B}\ }\textbf {\bibinfo {volume} {98}},\ \bibinfo {pages} {205136} (\bibinfo {year} {2018})}\BibitemShut {NoStop}%
\bibitem [{\citenamefont {Skinner}\ \emph {et~al.}(2019)\citenamefont {Skinner}, \citenamefont {Ruhman},\ and\ \citenamefont {Nahum}}]{skinner_nahum_prx2019}%
  \BibitemOpen
  \bibfield  {author} {\bibinfo {author} {\bibfnamefont {B.}~\bibnamefont {Skinner}}, \bibinfo {author} {\bibfnamefont {J.}~\bibnamefont {Ruhman}},\ and\ \bibinfo {author} {\bibfnamefont {A.}~\bibnamefont {Nahum}},\ }\bibfield  {title} {\bibinfo {title} {Measurement-induced phase transitions in the dynamics of entanglement},\ }\href {https://doi.org/10.1103/PhysRevX.9.031009} {\bibfield  {journal} {\bibinfo  {journal} {Phys. Rev. X}\ }\textbf {\bibinfo {volume} {9}},\ \bibinfo {pages} {031009} (\bibinfo {year} {2019})}\BibitemShut {NoStop}%
\bibitem [{\citenamefont {Li}\ \emph {et~al.}(2019)\citenamefont {Li}, \citenamefont {Chen},\ and\ \citenamefont {Fisher}}]{Li_Chen_Fisher}%
  \BibitemOpen
  \bibfield  {author} {\bibinfo {author} {\bibfnamefont {Y.}~\bibnamefont {Li}}, \bibinfo {author} {\bibfnamefont {X.}~\bibnamefont {Chen}},\ and\ \bibinfo {author} {\bibfnamefont {M.~P.~A.}\ \bibnamefont {Fisher}},\ }\bibfield  {title} {\bibinfo {title} {Measurement-driven entanglement transition in hybrid quantum circuits},\ }\href {https://doi.org/10.1103/PhysRevB.100.134306} {\bibfield  {journal} {\bibinfo  {journal} {Phys. Rev. B}\ }\textbf {\bibinfo {volume} {100}},\ \bibinfo {pages} {134306} (\bibinfo {year} {2019})}\BibitemShut {NoStop}%
\bibitem [{\citenamefont {Nielsen}\ and\ \citenamefont {Chuang}(2010)}]{Nielsen_Chuang_2010}%
  \BibitemOpen
  \bibfield  {author} {\bibinfo {author} {\bibfnamefont {M.~A.}\ \bibnamefont {Nielsen}}\ and\ \bibinfo {author} {\bibfnamefont {I.~L.}\ \bibnamefont {Chuang}},\ }\href@noop {} {\emph {\bibinfo {title} {Quantum Computation and Quantum Information: 10th Anniversary Edition}}}\ (\bibinfo  {publisher} {Cambridge University Press},\ \bibinfo {year} {2010})\BibitemShut {NoStop}%
\bibitem [{\citenamefont {Lewis-Swan}\ \emph {et~al.}(2019)\citenamefont {Lewis-Swan}, \citenamefont {Safavi-Naini}, \citenamefont {Kaufman},\ and\ \citenamefont {Rey}}]{dynamics_qi_nat.rev.phy2019}%
  \BibitemOpen
  \bibfield  {author} {\bibinfo {author} {\bibfnamefont {R.~J.}\ \bibnamefont {Lewis-Swan}}, \bibinfo {author} {\bibfnamefont {A.}~\bibnamefont {Safavi-Naini}}, \bibinfo {author} {\bibfnamefont {A.~M.}\ \bibnamefont {Kaufman}},\ and\ \bibinfo {author} {\bibfnamefont {A.~M.}\ \bibnamefont {Rey}},\ }\bibfield  {title} {\bibinfo {title} {Dynamics of quantum information},\ }\href {https://doi.org/10.1038/s42254-019-0090-y} {\bibfield  {journal} {\bibinfo  {journal} {Nature Reviews Physics}\ }\textbf {\bibinfo {volume} {1}},\ \bibinfo {pages} {627} (\bibinfo {year} {2019})}\BibitemShut {NoStop}%
\bibitem [{\citenamefont {Vidal}(2003)}]{vidal_mps_prl2003}%
  \BibitemOpen
  \bibfield  {author} {\bibinfo {author} {\bibfnamefont {G.}~\bibnamefont {Vidal}},\ }\bibfield  {title} {\bibinfo {title} {Efficient classical simulation of slightly entangled quantum computations},\ }\href {https://doi.org/10.1103/PhysRevLett.91.147902} {\bibfield  {journal} {\bibinfo  {journal} {Phys. Rev. Lett.}\ }\textbf {\bibinfo {volume} {91}},\ \bibinfo {pages} {147902} (\bibinfo {year} {2003})}\BibitemShut {NoStop}%
\bibitem [{\citenamefont {Verstraete}\ \emph {et~al.}(2004)\citenamefont {Verstraete}, \citenamefont {Garc\'{\i}a-Ripoll},\ and\ \citenamefont {Cirac}}]{verstraete_cirac_mps_prl2004}%
  \BibitemOpen
  \bibfield  {author} {\bibinfo {author} {\bibfnamefont {F.}~\bibnamefont {Verstraete}}, \bibinfo {author} {\bibfnamefont {J.~J.}\ \bibnamefont {Garc\'{\i}a-Ripoll}},\ and\ \bibinfo {author} {\bibfnamefont {J.~I.}\ \bibnamefont {Cirac}},\ }\bibfield  {title} {\bibinfo {title} {Matrix product density operators: Simulation of finite-temperature and dissipative systems},\ }\href {https://doi.org/10.1103/PhysRevLett.93.207204} {\bibfield  {journal} {\bibinfo  {journal} {Phys. Rev. Lett.}\ }\textbf {\bibinfo {volume} {93}},\ \bibinfo {pages} {207204} (\bibinfo {year} {2004})}\BibitemShut {NoStop}%
\bibitem [{\citenamefont {White}\ and\ \citenamefont {Feiguin}(2004)}]{white_feiguin_tebd_prl2004}%
  \BibitemOpen
  \bibfield  {author} {\bibinfo {author} {\bibfnamefont {S.~R.}\ \bibnamefont {White}}\ and\ \bibinfo {author} {\bibfnamefont {A.~E.}\ \bibnamefont {Feiguin}},\ }\bibfield  {title} {\bibinfo {title} {Real-time evolution using the density matrix renormalization group},\ }\href {https://doi.org/10.1103/PhysRevLett.93.076401} {\bibfield  {journal} {\bibinfo  {journal} {Phys. Rev. Lett.}\ }\textbf {\bibinfo {volume} {93}},\ \bibinfo {pages} {076401} (\bibinfo {year} {2004})}\BibitemShut {NoStop}%
\bibitem [{\citenamefont {Daley}\ \emph {et~al.}(2004)\citenamefont {Daley}, \citenamefont {Kollath}, \citenamefont {Schollwöck},\ and\ \citenamefont {Vidal}}]{daley_vidal_tebd2004}%
  \BibitemOpen
  \bibfield  {author} {\bibinfo {author} {\bibfnamefont {A.~J.}\ \bibnamefont {Daley}}, \bibinfo {author} {\bibfnamefont {C.}~\bibnamefont {Kollath}}, \bibinfo {author} {\bibfnamefont {U.}~\bibnamefont {Schollwöck}},\ and\ \bibinfo {author} {\bibfnamefont {G.}~\bibnamefont {Vidal}},\ }\bibfield  {title} {\bibinfo {title} {Time-dependent density-matrix renormalization-group using adaptive effective hilbert spaces},\ }\href {https://doi.org/10.1088/1742-5468/2004/04/P04005} {\bibfield  {journal} {\bibinfo  {journal} {Journal of Statistical Mechanics: Theory and Experiment}\ }\textbf {\bibinfo {volume} {2004}},\ \bibinfo {pages} {P04005} (\bibinfo {year} {2004})}\BibitemShut {NoStop}%
\bibitem [{\citenamefont {Eisert}\ \emph {et~al.}(2010)\citenamefont {Eisert}, \citenamefont {Cramer},\ and\ \citenamefont {Plenio}}]{Eisert_Plenio_RMP2010}%
  \BibitemOpen
  \bibfield  {author} {\bibinfo {author} {\bibfnamefont {J.}~\bibnamefont {Eisert}}, \bibinfo {author} {\bibfnamefont {M.}~\bibnamefont {Cramer}},\ and\ \bibinfo {author} {\bibfnamefont {M.~B.}\ \bibnamefont {Plenio}},\ }\bibfield  {title} {\bibinfo {title} {Colloquium: Area laws for the entanglement entropy},\ }\href {https://doi.org/10.1103/RevModPhys.82.277} {\bibfield  {journal} {\bibinfo  {journal} {Rev. Mod. Phys.}\ }\textbf {\bibinfo {volume} {82}},\ \bibinfo {pages} {277} (\bibinfo {year} {2010})}\BibitemShut {NoStop}%
\bibitem [{\citenamefont {Nahum}\ \emph {et~al.}(2017)\citenamefont {Nahum}, \citenamefont {Ruhman}, \citenamefont {Vijay},\ and\ \citenamefont {Haah}}]{Nahum_Haah_entanglementgrowth_2016}%
  \BibitemOpen
  \bibfield  {author} {\bibinfo {author} {\bibfnamefont {A.}~\bibnamefont {Nahum}}, \bibinfo {author} {\bibfnamefont {J.}~\bibnamefont {Ruhman}}, \bibinfo {author} {\bibfnamefont {S.}~\bibnamefont {Vijay}},\ and\ \bibinfo {author} {\bibfnamefont {J.}~\bibnamefont {Haah}},\ }\bibfield  {title} {\bibinfo {title} {Quantum entanglement growth under random unitary dynamics},\ }\href {https://doi.org/10.1103/PhysRevX.7.031016} {\bibfield  {journal} {\bibinfo  {journal} {Phys. Rev. X}\ }\textbf {\bibinfo {volume} {7}},\ \bibinfo {pages} {031016} (\bibinfo {year} {2017})}\BibitemShut {NoStop}%
\bibitem [{\citenamefont {Jonay}\ \emph {et~al.}(2018)\citenamefont {Jonay}, \citenamefont {Huse},\ and\ \citenamefont {Nahum}}]{Jonay2018}%
  \BibitemOpen
  \bibfield  {author} {\bibinfo {author} {\bibfnamefont {C.}~\bibnamefont {Jonay}}, \bibinfo {author} {\bibfnamefont {D.~A.}\ \bibnamefont {Huse}},\ and\ \bibinfo {author} {\bibfnamefont {A.}~\bibnamefont {Nahum}},\ }\href {http://arxiv.org/abs/1803.00089} {\bibinfo {title} {{Coarse-grained dynamics of operator and state entanglement}}} (\bibinfo {year} {2018}),\ \Eprint {https://arxiv.org/abs/1803.00089} {arXiv:1803.00089} \BibitemShut {NoStop}%
\bibitem [{\citenamefont {Zhou}\ and\ \citenamefont {Nahum}(2019)}]{Zhou2019}%
  \BibitemOpen
  \bibfield  {author} {\bibinfo {author} {\bibfnamefont {T.}~\bibnamefont {Zhou}}\ and\ \bibinfo {author} {\bibfnamefont {A.}~\bibnamefont {Nahum}},\ }\bibfield  {title} {\bibinfo {title} {{Emergent statistical mechanics of entanglement in random unitary circuits}},\ }\href {https://doi.org/10.1103/PhysRevB.99.174205} {\bibfield  {journal} {\bibinfo  {journal} {Physical Review B}\ }\textbf {\bibinfo {volume} {99}},\ \bibinfo {pages} {174205} (\bibinfo {year} {2019})},\ \Eprint {https://arxiv.org/abs/1804.09737} {1804.09737} \BibitemShut {NoStop}%
\bibitem [{\citenamefont {Zhou}\ and\ \citenamefont {Nahum}(2020)}]{Zhou2020}%
  \BibitemOpen
  \bibfield  {author} {\bibinfo {author} {\bibfnamefont {T.}~\bibnamefont {Zhou}}\ and\ \bibinfo {author} {\bibfnamefont {A.}~\bibnamefont {Nahum}},\ }\bibfield  {title} {\bibinfo {title} {Entanglement membrane in chaotic many-body systems},\ }\href {https://doi.org/10.1103/PhysRevX.10.031066} {\bibfield  {journal} {\bibinfo  {journal} {Phys. Rev. X}\ }\textbf {\bibinfo {volume} {10}},\ \bibinfo {pages} {031066} (\bibinfo {year} {2020})}\BibitemShut {NoStop}%
\bibitem [{\citenamefont {Sierant}\ \emph {et~al.}(2023)\citenamefont {Sierant}, \citenamefont {Schir\`o}, \citenamefont {Lewenstein},\ and\ \citenamefont {Turkeshi}}]{sierant_turkeshi_membrane2023}%
  \BibitemOpen
  \bibfield  {author} {\bibinfo {author} {\bibfnamefont {P.}~\bibnamefont {Sierant}}, \bibinfo {author} {\bibfnamefont {M.}~\bibnamefont {Schir\`o}}, \bibinfo {author} {\bibfnamefont {M.}~\bibnamefont {Lewenstein}},\ and\ \bibinfo {author} {\bibfnamefont {X.}~\bibnamefont {Turkeshi}},\ }\bibfield  {title} {\bibinfo {title} {Entanglement growth and minimal membranes in ($d+1$) random unitary circuits},\ }\href {https://doi.org/10.1103/PhysRevLett.131.230403} {\bibfield  {journal} {\bibinfo  {journal} {Phys. Rev. Lett.}\ }\textbf {\bibinfo {volume} {131}},\ \bibinfo {pages} {230403} (\bibinfo {year} {2023})}\BibitemShut {NoStop}%
\bibitem [{\citenamefont {Ryu}\ and\ \citenamefont {Takayanagi}(2006)}]{Ryu_Takayanagi_2006}%
  \BibitemOpen
  \bibfield  {author} {\bibinfo {author} {\bibfnamefont {S.}~\bibnamefont {Ryu}}\ and\ \bibinfo {author} {\bibfnamefont {T.}~\bibnamefont {Takayanagi}},\ }\bibfield  {title} {\bibinfo {title} {Holographic derivation of entanglement entropy from the anti--de sitter space/conformal field theory correspondence},\ }\href {https://doi.org/10.1103/PhysRevLett.96.181602} {\bibfield  {journal} {\bibinfo  {journal} {Phys. Rev. Lett.}\ }\textbf {\bibinfo {volume} {96}},\ \bibinfo {pages} {181602} (\bibinfo {year} {2006})}\BibitemShut {NoStop}%
\bibitem [{\citenamefont {Fisher}\ \emph {et~al.}(2023)\citenamefont {Fisher}, \citenamefont {Khemani}, \citenamefont {Nahum},\ and\ \citenamefont {Vijay}}]{Random_Quantum_Circuits_review2023}%
  \BibitemOpen
  \bibfield  {author} {\bibinfo {author} {\bibfnamefont {M.~P.}\ \bibnamefont {Fisher}}, \bibinfo {author} {\bibfnamefont {V.}~\bibnamefont {Khemani}}, \bibinfo {author} {\bibfnamefont {A.}~\bibnamefont {Nahum}},\ and\ \bibinfo {author} {\bibfnamefont {S.}~\bibnamefont {Vijay}},\ }\bibfield  {title} {\bibinfo {title} {Random quantum circuits},\ }\href {https://doi.org/https://doi.org/10.1146/annurev-conmatphys-031720-030658} {\bibfield  {journal} {\bibinfo  {journal} {Annual Review of Condensed Matter Physics}\ }\textbf {\bibinfo {volume} {14}},\ \bibinfo {pages} {335} (\bibinfo {year} {2023})}\BibitemShut {NoStop}%
\bibitem [{\citenamefont {Foligno}\ \emph {et~al.}(2024)\citenamefont {Foligno}, \citenamefont {Kos},\ and\ \citenamefont {Bertini}}]{Foligno_Kos_Bertini_prl2024}%
  \BibitemOpen
  \bibfield  {author} {\bibinfo {author} {\bibfnamefont {A.}~\bibnamefont {Foligno}}, \bibinfo {author} {\bibfnamefont {P.}~\bibnamefont {Kos}},\ and\ \bibinfo {author} {\bibfnamefont {B.}~\bibnamefont {Bertini}},\ }\bibfield  {title} {\bibinfo {title} {Quantum information spreading in generalized dual-unitary circuits},\ }\href {https://doi.org/10.1103/PhysRevLett.132.250402} {\bibfield  {journal} {\bibinfo  {journal} {Phys. Rev. Lett.}\ }\textbf {\bibinfo {volume} {132}},\ \bibinfo {pages} {250402} (\bibinfo {year} {2024})}\BibitemShut {NoStop}%
\bibitem [{\citenamefont {Rampp}\ \emph {et~al.}(2024)\citenamefont {Rampp}, \citenamefont {Rather},\ and\ \citenamefont {Claeys}}]{Rampp_Claeys_PRR2024}%
  \BibitemOpen
  \bibfield  {author} {\bibinfo {author} {\bibfnamefont {M.~A.}\ \bibnamefont {Rampp}}, \bibinfo {author} {\bibfnamefont {S.~A.}\ \bibnamefont {Rather}},\ and\ \bibinfo {author} {\bibfnamefont {P.~W.}\ \bibnamefont {Claeys}},\ }\bibfield  {title} {\bibinfo {title} {Entanglement membrane in exactly solvable lattice models},\ }\href {https://doi.org/10.1103/PhysRevResearch.6.033271} {\bibfield  {journal} {\bibinfo  {journal} {Phys. Rev. Res.}\ }\textbf {\bibinfo {volume} {6}},\ \bibinfo {pages} {033271} (\bibinfo {year} {2024})}\BibitemShut {NoStop}%
\bibitem [{\citenamefont {Claeys}\ and\ \citenamefont {Lamacraft}(2024)}]{Claeys_lamacraft_2024}%
  \BibitemOpen
  \bibfield  {author} {\bibinfo {author} {\bibfnamefont {P.~W.}\ \bibnamefont {Claeys}}\ and\ \bibinfo {author} {\bibfnamefont {A.}~\bibnamefont {Lamacraft}},\ }\bibfield  {title} {\bibinfo {title} {Operator dynamics and entanglement in space-time dual hadamard lattices},\ }\href {https://doi.org/10.1088/1751-8121/ad776a} {\bibfield  {journal} {\bibinfo  {journal} {Journal of Physics A: Mathematical and Theoretical}\ }\textbf {\bibinfo {volume} {57}},\ \bibinfo {pages} {405301} (\bibinfo {year} {2024})}\BibitemShut {NoStop}%
\bibitem [{\citenamefont {Liu}\ and\ \citenamefont {Ho}(2025)}]{liu_ho_prr2025}%
  \BibitemOpen
  \bibfield  {author} {\bibinfo {author} {\bibfnamefont {C.}~\bibnamefont {Liu}}\ and\ \bibinfo {author} {\bibfnamefont {W.~W.}\ \bibnamefont {Ho}},\ }\bibfield  {title} {\bibinfo {title} {Solvable entanglement dynamics in quantum circuits with generalized space-time duality},\ }\href {https://doi.org/10.1103/PhysRevResearch.7.L012011} {\bibfield  {journal} {\bibinfo  {journal} {Phys. Rev. Res.}\ }\textbf {\bibinfo {volume} {7}},\ \bibinfo {pages} {L012011} (\bibinfo {year} {2025})}\BibitemShut {NoStop}%
\bibitem [{\citenamefont {Vasseur}\ \emph {et~al.}(2019)\citenamefont {Vasseur}, \citenamefont {Potter}, \citenamefont {You},\ and\ \citenamefont {Ludwig}}]{vasseur_ludwig_PRB2019}%
  \BibitemOpen
  \bibfield  {author} {\bibinfo {author} {\bibfnamefont {R.}~\bibnamefont {Vasseur}}, \bibinfo {author} {\bibfnamefont {A.~C.}\ \bibnamefont {Potter}}, \bibinfo {author} {\bibfnamefont {Y.-Z.}\ \bibnamefont {You}},\ and\ \bibinfo {author} {\bibfnamefont {A.~W.~W.}\ \bibnamefont {Ludwig}},\ }\bibfield  {title} {\bibinfo {title} {Entanglement transitions from holographic random tensor networks},\ }\href {https://doi.org/10.1103/PhysRevB.100.134203} {\bibfield  {journal} {\bibinfo  {journal} {Phys. Rev. B}\ }\textbf {\bibinfo {volume} {100}},\ \bibinfo {pages} {134203} (\bibinfo {year} {2019})}\BibitemShut {NoStop}%
\bibitem [{\citenamefont {Mezei}(2018)}]{mezei_prd2018}%
  \BibitemOpen
  \bibfield  {author} {\bibinfo {author} {\bibfnamefont {M.}~\bibnamefont {Mezei}},\ }\bibfield  {title} {\bibinfo {title} {Membrane theory of entanglement dynamics from holography},\ }\href {https://doi.org/10.1103/PhysRevD.98.106025} {\bibfield  {journal} {\bibinfo  {journal} {Phys. Rev. D}\ }\textbf {\bibinfo {volume} {98}},\ \bibinfo {pages} {106025} (\bibinfo {year} {2018})}\BibitemShut {NoStop}%
\bibitem [{\citenamefont {Mezei}\ and\ \citenamefont {Virrueta}(2020)}]{Mezei_jhep2020}%
  \BibitemOpen
  \bibfield  {author} {\bibinfo {author} {\bibfnamefont {M.}~\bibnamefont {Mezei}}\ and\ \bibinfo {author} {\bibfnamefont {J.}~\bibnamefont {Virrueta}},\ }\bibfield  {title} {\bibinfo {title} {Exploring the membrane theory of entanglement dynamics},\ }\href {https://doi.org/10.1007/JHEP02(2020)013} {\bibfield  {journal} {\bibinfo  {journal} {Journal of High Energy Physics}\ }\textbf {\bibinfo {volume} {2020}},\ \bibinfo {pages} {13} (\bibinfo {year} {2020})}\BibitemShut {NoStop}%
\bibitem [{\citenamefont {Jiang}\ \emph {et~al.}(2024)\citenamefont {Jiang}, \citenamefont {Mezei},\ and\ \citenamefont {Virrueta}}]{jiang2024}%
  \BibitemOpen
  \bibfield  {author} {\bibinfo {author} {\bibfnamefont {H.}~\bibnamefont {Jiang}}, \bibinfo {author} {\bibfnamefont {M.}~\bibnamefont {Mezei}},\ and\ \bibinfo {author} {\bibfnamefont {J.}~\bibnamefont {Virrueta}},\ }\href {https://arxiv.org/abs/2411.16542} {\bibinfo {title} {The entanglement membrane in 2d cft: reflected entropy, rg flow, and information velocity}} (\bibinfo {year} {2024}),\ \Eprint {https://arxiv.org/abs/2411.16542} {arXiv:2411.16542 [hep-th]} \BibitemShut {NoStop}%
\bibitem [{\citenamefont {Ippoliti}\ \emph {et~al.}(2022)\citenamefont {Ippoliti}, \citenamefont {Rakovszky},\ and\ \citenamefont {Khemani}}]{ippoliti_rakovszky_khemani_prx2022}%
  \BibitemOpen
  \bibfield  {author} {\bibinfo {author} {\bibfnamefont {M.}~\bibnamefont {Ippoliti}}, \bibinfo {author} {\bibfnamefont {T.}~\bibnamefont {Rakovszky}},\ and\ \bibinfo {author} {\bibfnamefont {V.}~\bibnamefont {Khemani}},\ }\bibfield  {title} {\bibinfo {title} {Fractal, logarithmic, and volume-law entangled nonthermal steady states via spacetime duality},\ }\href {https://doi.org/10.1103/PhysRevX.12.011045} {\bibfield  {journal} {\bibinfo  {journal} {Phys. Rev. X}\ }\textbf {\bibinfo {volume} {12}},\ \bibinfo {pages} {011045} (\bibinfo {year} {2022})}\BibitemShut {NoStop}%
\bibitem [{\citenamefont {Lovas}\ \emph {et~al.}(2024)\citenamefont {Lovas}, \citenamefont {Agrawal},\ and\ \citenamefont {Vijay}}]{Lovas_vijay_prxq2024}%
  \BibitemOpen
  \bibfield  {author} {\bibinfo {author} {\bibfnamefont {I.}~\bibnamefont {Lovas}}, \bibinfo {author} {\bibfnamefont {U.}~\bibnamefont {Agrawal}},\ and\ \bibinfo {author} {\bibfnamefont {S.}~\bibnamefont {Vijay}},\ }\bibfield  {title} {\bibinfo {title} {Quantum coding transitions in the presence of boundary dissipation},\ }\href {https://doi.org/10.1103/PRXQuantum.5.030327} {\bibfield  {journal} {\bibinfo  {journal} {PRX Quantum}\ }\textbf {\bibinfo {volume} {5}},\ \bibinfo {pages} {030327} (\bibinfo {year} {2024})}\BibitemShut {NoStop}%
\bibitem [{\citenamefont {Jian}\ \emph {et~al.}(2020)\citenamefont {Jian}, \citenamefont {You}, \citenamefont {Vasseur},\ and\ \citenamefont {Ludwig}}]{jian_ludwig_prb2020}%
  \BibitemOpen
  \bibfield  {author} {\bibinfo {author} {\bibfnamefont {C.-M.}\ \bibnamefont {Jian}}, \bibinfo {author} {\bibfnamefont {Y.-Z.}\ \bibnamefont {You}}, \bibinfo {author} {\bibfnamefont {R.}~\bibnamefont {Vasseur}},\ and\ \bibinfo {author} {\bibfnamefont {A.~W.~W.}\ \bibnamefont {Ludwig}},\ }\bibfield  {title} {\bibinfo {title} {Measurement-induced criticality in random quantum circuits},\ }\href {https://doi.org/10.1103/PhysRevB.101.104302} {\bibfield  {journal} {\bibinfo  {journal} {Phys. Rev. B}\ }\textbf {\bibinfo {volume} {101}},\ \bibinfo {pages} {104302} (\bibinfo {year} {2020})}\BibitemShut {NoStop}%
\bibitem [{\citenamefont {Bao}\ \emph {et~al.}(2020)\citenamefont {Bao}, \citenamefont {Choi},\ and\ \citenamefont {Altman}}]{bao_choi_altman_prb2020}%
  \BibitemOpen
  \bibfield  {author} {\bibinfo {author} {\bibfnamefont {Y.}~\bibnamefont {Bao}}, \bibinfo {author} {\bibfnamefont {S.}~\bibnamefont {Choi}},\ and\ \bibinfo {author} {\bibfnamefont {E.}~\bibnamefont {Altman}},\ }\bibfield  {title} {\bibinfo {title} {Theory of the phase transition in random unitary circuits with measurements},\ }\href {https://doi.org/10.1103/PhysRevB.101.104301} {\bibfield  {journal} {\bibinfo  {journal} {Phys. Rev. B}\ }\textbf {\bibinfo {volume} {101}},\ \bibinfo {pages} {104301} (\bibinfo {year} {2020})}\BibitemShut {NoStop}%
\bibitem [{\citenamefont {Li}\ \emph {et~al.}(2023)\citenamefont {Li}, \citenamefont {Vijay},\ and\ \citenamefont {Fisher}}]{Li_Vijay_Fisher_DPRE2023}%
  \BibitemOpen
  \bibfield  {author} {\bibinfo {author} {\bibfnamefont {Y.}~\bibnamefont {Li}}, \bibinfo {author} {\bibfnamefont {S.}~\bibnamefont {Vijay}},\ and\ \bibinfo {author} {\bibfnamefont {M.~P.}\ \bibnamefont {Fisher}},\ }\bibfield  {title} {\bibinfo {title} {Entanglement domain walls in monitored quantum circuits and the directed polymer in a random environment},\ }\href {https://doi.org/10.1103/PRXQuantum.4.010331} {\bibfield  {journal} {\bibinfo  {journal} {PRX Quantum}\ }\textbf {\bibinfo {volume} {4}},\ \bibinfo {pages} {010331} (\bibinfo {year} {2023})}\BibitemShut {NoStop}%
\bibitem [{\citenamefont {Sommers}\ \emph {et~al.}(2024)\citenamefont {Sommers}, \citenamefont {Gopalakrishnan}, \citenamefont {Gullans},\ and\ \citenamefont {Huse}}]{Sommers_Huse_2024}%
  \BibitemOpen
  \bibfield  {author} {\bibinfo {author} {\bibfnamefont {G.~M.}\ \bibnamefont {Sommers}}, \bibinfo {author} {\bibfnamefont {S.}~\bibnamefont {Gopalakrishnan}}, \bibinfo {author} {\bibfnamefont {M.~J.}\ \bibnamefont {Gullans}},\ and\ \bibinfo {author} {\bibfnamefont {D.~A.}\ \bibnamefont {Huse}},\ }\bibfield  {title} {\bibinfo {title} {Zero-temperature entanglement membranes in quantum circuits},\ }\href {https://doi.org/10.1103/PhysRevB.110.064311} {\bibfield  {journal} {\bibinfo  {journal} {Phys. Rev. B}\ }\textbf {\bibinfo {volume} {110}},\ \bibinfo {pages} {064311} (\bibinfo {year} {2024})}\BibitemShut {NoStop}%
\bibitem [{\citenamefont {Huse}\ and\ \citenamefont {Henley}(1985)}]{Huse_Henley_1985}%
  \BibitemOpen
  \bibfield  {author} {\bibinfo {author} {\bibfnamefont {D.~A.}\ \bibnamefont {Huse}}\ and\ \bibinfo {author} {\bibfnamefont {C.~L.}\ \bibnamefont {Henley}},\ }\bibfield  {title} {\bibinfo {title} {Pinning and roughening of domain walls in ising systems due to random impurities},\ }\href {https://doi.org/10.1103/PhysRevLett.54.2708} {\bibfield  {journal} {\bibinfo  {journal} {Phys. Rev. Lett.}\ }\textbf {\bibinfo {volume} {54}},\ \bibinfo {pages} {2708} (\bibinfo {year} {1985})}\BibitemShut {NoStop}%
\bibitem [{\citenamefont {Huse}\ \emph {et~al.}(1985)\citenamefont {Huse}, \citenamefont {Henley},\ and\ \citenamefont {Fisher}}]{Huse_Henley_Fisher_respond_1985}%
  \BibitemOpen
  \bibfield  {author} {\bibinfo {author} {\bibfnamefont {D.~A.}\ \bibnamefont {Huse}}, \bibinfo {author} {\bibfnamefont {C.~L.}\ \bibnamefont {Henley}},\ and\ \bibinfo {author} {\bibfnamefont {D.~S.}\ \bibnamefont {Fisher}},\ }\bibfield  {title} {\bibinfo {title} {Huse, henley, and fisher respond},\ }\href {https://doi.org/10.1103/PhysRevLett.55.2924} {\bibfield  {journal} {\bibinfo  {journal} {Phys. Rev. Lett.}\ }\textbf {\bibinfo {volume} {55}},\ \bibinfo {pages} {2924} (\bibinfo {year} {1985})}\BibitemShut {NoStop}%
\bibitem [{\citenamefont {Halpin-Healy}(1990)}]{Halpin-Healy_FRG_1990}%
  \BibitemOpen
  \bibfield  {author} {\bibinfo {author} {\bibfnamefont {T.}~\bibnamefont {Halpin-Healy}},\ }\bibfield  {title} {\bibinfo {title} {Disorder-induced roughening of diverse manifolds},\ }\href {https://doi.org/10.1103/PhysRevA.42.711} {\bibfield  {journal} {\bibinfo  {journal} {Phys. Rev. A}\ }\textbf {\bibinfo {volume} {42}},\ \bibinfo {pages} {711} (\bibinfo {year} {1990})}\BibitemShut {NoStop}%
\bibitem [{\citenamefont {Emig}\ and\ \citenamefont {Nattermann}(1998)}]{Emig_Nattermann_1998_FRGprl}%
  \BibitemOpen
  \bibfield  {author} {\bibinfo {author} {\bibfnamefont {T.}~\bibnamefont {Emig}}\ and\ \bibinfo {author} {\bibfnamefont {T.}~\bibnamefont {Nattermann}},\ }\bibfield  {title} {\bibinfo {title} {Roughening transition of interfaces in disordered systems},\ }\href {https://doi.org/10.1103/PhysRevLett.81.1469} {\bibfield  {journal} {\bibinfo  {journal} {Phys. Rev. Lett.}\ }\textbf {\bibinfo {volume} {81}},\ \bibinfo {pages} {1469} (\bibinfo {year} {1998})}\BibitemShut {NoStop}%
\bibitem [{\citenamefont {Emig}\ and\ \citenamefont {Nattermann}(1999)}]{Emig_Nattermann_1998_FRGlong}%
  \BibitemOpen
  \bibfield  {author} {\bibinfo {author} {\bibfnamefont {T.}~\bibnamefont {Emig}}\ and\ \bibinfo {author} {\bibfnamefont {T.}~\bibnamefont {Nattermann}},\ }\bibfield  {title} {\bibinfo {title} {Disorder driven roughening transitions of elastic manifolds and periodic elastic media},\ }\href {https://doi.org/10.1007/s100510050720} {\bibfield  {journal} {\bibinfo  {journal} {The European Physical Journal B - Condensed Matter and Complex Systems}\ }\textbf {\bibinfo {volume} {8}},\ \bibinfo {pages} {525} (\bibinfo {year} {1999})}\BibitemShut {NoStop}%
\bibitem [{\citenamefont {Noh}\ and\ \citenamefont {Rieger}(2002)}]{Noh_Rieger_2002_numerics}%
  \BibitemOpen
  \bibfield  {author} {\bibinfo {author} {\bibfnamefont {J.~D.}\ \bibnamefont {Noh}}\ and\ \bibinfo {author} {\bibfnamefont {H.}~\bibnamefont {Rieger}},\ }\bibfield  {title} {\bibinfo {title} {Numerical study of the disorder-driven roughening transition in an elastic manifold in a periodic potential},\ }\href {https://doi.org/10.1103/PhysRevE.66.036117} {\bibfield  {journal} {\bibinfo  {journal} {Phys. Rev. E}\ }\textbf {\bibinfo {volume} {66}},\ \bibinfo {pages} {036117} (\bibinfo {year} {2002})}\BibitemShut {NoStop}%
\bibitem [{\citenamefont {Bassan}\ \emph {et~al.}(2023)\citenamefont {Bassan}, \citenamefont {Gilboa},\ and\ \citenamefont {Peled}}]{Peled_mathproof2023}%
  \BibitemOpen
  \bibfield  {author} {\bibinfo {author} {\bibfnamefont {M.}~\bibnamefont {Bassan}}, \bibinfo {author} {\bibfnamefont {S.}~\bibnamefont {Gilboa}},\ and\ \bibinfo {author} {\bibfnamefont {R.}~\bibnamefont {Peled}},\ }\href {https://arxiv.org/abs/2309.06437} {\bibinfo {title} {Non-constant ground configurations in the disordered ferromagnet}} (\bibinfo {year} {2023}),\ \Eprint {https://arxiv.org/abs/2309.06437} {arXiv:2309.06437 [math-ph]} \BibitemShut {NoStop}%
\bibitem [{\citenamefont {BOVIER}\ and\ \citenamefont {K\"{U}LSKE}(1994)}]{Bovier-Külske_3d}%
  \BibitemOpen
  \bibfield  {author} {\bibinfo {author} {\bibfnamefont {A.}~\bibnamefont {BOVIER}}\ and\ \bibinfo {author} {\bibfnamefont {C.}~\bibnamefont {K\"{U}LSKE}},\ }\bibfield  {title} {\bibinfo {title} {A rigorous renormalization group method for interfaces in random media},\ }\href {https://doi.org/10.1142/S0129055X94000171} {\bibfield  {journal} {\bibinfo  {journal} {Reviews in Mathematical Physics}\ }\textbf {\bibinfo {volume} {06}},\ \bibinfo {pages} {413} (\bibinfo {year} {1994})},\ \Eprint {https://arxiv.org/abs/https://doi.org/10.1142/S0129055X94000171} {https://doi.org/10.1142/S0129055X94000171} \BibitemShut {NoStop}%
\bibitem [{\citenamefont {Bovier}\ and\ \citenamefont {K{\"u}lske}(1996)}]{Bovier-Külske_2d}%
  \BibitemOpen
  \bibfield  {author} {\bibinfo {author} {\bibfnamefont {A.}~\bibnamefont {Bovier}}\ and\ \bibinfo {author} {\bibfnamefont {C.}~\bibnamefont {K{\"u}lske}},\ }\bibfield  {title} {\bibinfo {title} {There are no nice interfaces in (2+1)-dimensional sos models in random media},\ }\href {https://doi.org/10.1007/BF02183747} {\bibfield  {journal} {\bibinfo  {journal} {Journal of Statistical Physics}\ }\textbf {\bibinfo {volume} {83}},\ \bibinfo {pages} {751} (\bibinfo {year} {1996})}\BibitemShut {NoStop}%
\bibitem [{\citenamefont {Dario}\ \emph {et~al.}(2023)\citenamefont {Dario}, \citenamefont {Harel},\ and\ \citenamefont {Peled}}]{peled_2023_math_randomsurface}%
  \BibitemOpen
  \bibfield  {author} {\bibinfo {author} {\bibfnamefont {P.}~\bibnamefont {Dario}}, \bibinfo {author} {\bibfnamefont {M.}~\bibnamefont {Harel}},\ and\ \bibinfo {author} {\bibfnamefont {R.}~\bibnamefont {Peled}},\ }\bibfield  {title} {\bibinfo {title} {Random-field random surfaces},\ }\href {https://doi.org/10.1007/s00440-022-01179-0} {\bibfield  {journal} {\bibinfo  {journal} {Probability Theory and Related Fields}\ }\textbf {\bibinfo {volume} {186}},\ \bibinfo {pages} {91} (\bibinfo {year} {2023})}\BibitemShut {NoStop}%
\bibitem [{\citenamefont {Sachdev}(2011)}]{Sachdev_2011}%
  \BibitemOpen
  \bibfield  {author} {\bibinfo {author} {\bibfnamefont {S.}~\bibnamefont {Sachdev}},\ }\href@noop {} {\emph {\bibinfo {title} {Quantum Phase Transitions}}},\ \bibinfo {edition} {2nd}\ ed.\ (\bibinfo  {publisher} {Cambridge University Press},\ \bibinfo {year} {2011})\BibitemShut {NoStop}%
\bibitem [{\citenamefont {Feldmeier}\ and\ \citenamefont {Knap}(2021)}]{Feldmeier2021}%
  \BibitemOpen
  \bibfield  {author} {\bibinfo {author} {\bibfnamefont {J.}~\bibnamefont {Feldmeier}}\ and\ \bibinfo {author} {\bibfnamefont {M.}~\bibnamefont {Knap}},\ }\bibfield  {title} {\bibinfo {title} {{Critically Slow Operator Dynamics in Constrained Many-Body Systems}},\ }\href {https://doi.org/10.1103/PhysRevLett.127.235301} {\bibfield  {journal} {\bibinfo  {journal} {Physical Review Letters}\ }\textbf {\bibinfo {volume} {127}},\ \bibinfo {pages} {235301} (\bibinfo {year} {2021})},\ \Eprint {https://arxiv.org/abs/2106.05292} {2106.05292} \BibitemShut {NoStop}%
\bibitem [{Note1()}]{Note1}%
  \BibitemOpen
  \bibinfo {note} {We prepare a near-maximally entangled state by acting with a global random unitary Clifford circuit on the product state $|0\rangle ^{\otimes 2L^3}$. The resulting state is a Page-like state.}\BibitemShut {Stop}%
\bibitem [{\citenamefont {Sommers}\ \emph {et~al.}(2023)\citenamefont {Sommers}, \citenamefont {Huse},\ and\ \citenamefont {Gullans}}]{Sommers_Huse_Gullans_2023}%
  \BibitemOpen
  \bibfield  {author} {\bibinfo {author} {\bibfnamefont {G.~M.}\ \bibnamefont {Sommers}}, \bibinfo {author} {\bibfnamefont {D.~A.}\ \bibnamefont {Huse}},\ and\ \bibinfo {author} {\bibfnamefont {M.~J.}\ \bibnamefont {Gullans}},\ }\bibfield  {title} {\bibinfo {title} {Crystalline quantum circuits},\ }\href {https://doi.org/10.1103/PRXQuantum.4.030313} {\bibfield  {journal} {\bibinfo  {journal} {PRX Quantum}\ }\textbf {\bibinfo {volume} {4}},\ \bibinfo {pages} {030313} (\bibinfo {year} {2023})}\BibitemShut {NoStop}%
\bibitem [{Note2()}]{Note2}%
  \BibitemOpen
  \bibinfo {note} {This is why we do not use the simple hypercubic lattice with time running along $\protect \hat {t} = (1,1,1,1)$; such geometries typically lead to infinite butterfly velocity, which can artificially suppress entanglement growth~\cite {Sommers_Huse_2024}. We avoid this fine-tuned limit to focus on the roughening physics.}\BibitemShut {Stop}%
\bibitem [{sup()}]{supp_mat}%
  \BibitemOpen
  \href@noop {} {}\bibinfo {note} {See the Supplemental Material for additional details (see also references \cite{Nahum2017op, Bertini2019, Piroli2019, Fisher_1984, Fisher_5-epsilon_1986, Peled2401, Aizenman_Wehr_prl1989_ImryMa, Aizenman_Wehr_ImryMa, Wiese2022, Alava_Duxbury_2+1rough1996, Seppala_Duxbury_2+1rougheningtransition2001, VandeWetering2020, Peled2504, Imry_Ma_prl1975, grinstein_ma_1983, villain_prl1984} therein).}\BibitemShut {Stop}%
\bibitem [{Note3()}]{Note3}%
  \BibitemOpen
  \bibinfo {note} {One could try using an ancilla as a probe to detect some aspects of the membrane geometry as mentioned in \cite {Sommers_Huse_2024}.}\BibitemShut {Stop}%
\bibitem [{\citenamefont {Kawashima}\ and\ \citenamefont {Ito}(1993)}]{scaling}%
  \BibitemOpen
  \bibfield  {author} {\bibinfo {author} {\bibfnamefont {N.}~\bibnamefont {Kawashima}}\ and\ \bibinfo {author} {\bibfnamefont {N.}~\bibnamefont {Ito}},\ }\bibfield  {title} {\bibinfo {title} {Critical behavior of the three-dimensional ±j model in a magnetic field},\ }\href {https://doi.org/10.1143/JPSJ.62.435} {\bibfield  {journal} {\bibinfo  {journal} {Journal of the Physical Society of Japan}\ }\textbf {\bibinfo {volume} {62}},\ \bibinfo {pages} {435} (\bibinfo {year} {1993})},\ \Eprint {https://arxiv.org/abs/https://doi.org/10.1143/JPSJ.62.435} {https://doi.org/10.1143/JPSJ.62.435} \BibitemShut {NoStop}%
\bibitem [{\citenamefont {Dennis}\ \emph {et~al.}(2002)\citenamefont {Dennis}, \citenamefont {Kitaev}, \citenamefont {Landahl},\ and\ \citenamefont {Preskill}}]{dennis2002}%
  \BibitemOpen
  \bibfield  {author} {\bibinfo {author} {\bibfnamefont {E.}~\bibnamefont {Dennis}}, \bibinfo {author} {\bibfnamefont {A.}~\bibnamefont {Kitaev}}, \bibinfo {author} {\bibfnamefont {A.}~\bibnamefont {Landahl}},\ and\ \bibinfo {author} {\bibfnamefont {J.}~\bibnamefont {Preskill}},\ }\bibfield  {title} {\bibinfo {title} {Topological quantum memory},\ }\href {https://doi.org/10.1063/1.1499754} {\bibfield  {journal} {\bibinfo  {journal} {Journal of Mathematical Physics}\ }\textbf {\bibinfo {volume} {43}},\ \bibinfo {pages} {4452} (\bibinfo {year} {2002})}\BibitemShut {NoStop}%
\bibitem [{\citenamefont {Breuckmann}\ \emph {et~al.}(2017)\citenamefont {Breuckmann}, \citenamefont {Duivenvoorden}, \citenamefont {Michels},\ and\ \citenamefont {Terhal}}]{Breuckmann2017}%
  \BibitemOpen
  \bibfield  {author} {\bibinfo {author} {\bibfnamefont {N.~P.}\ \bibnamefont {Breuckmann}}, \bibinfo {author} {\bibfnamefont {K.}~\bibnamefont {Duivenvoorden}}, \bibinfo {author} {\bibfnamefont {D.}~\bibnamefont {Michels}},\ and\ \bibinfo {author} {\bibfnamefont {B.~M.}\ \bibnamefont {Terhal}},\ }\bibfield  {title} {\bibinfo {title} {Local decoders for the 2d and 4d toric code},\ }\href {https://www.scopus.com/inward/record.uri?eid=2-s2.0-85014481588&partnerID=40&md5=a3a4a7691e4a2d3887dc962f7fa8bdbf} {\bibfield  {journal} {\bibinfo  {journal} {Quantum Information and Computation}\ }\textbf {\bibinfo {volume} {17}},\ \bibinfo {pages} {181 – 208} (\bibinfo {year} {2017})}\BibitemShut {NoStop}%
\bibitem [{\citenamefont {Alicki}\ \emph {et~al.}(2010)\citenamefont {Alicki}, \citenamefont {Horodecki}, \citenamefont {Horodecki},\ and\ \citenamefont {Horodecki}}]{horodecki_4dtoric2010}%
  \BibitemOpen
  \bibfield  {author} {\bibinfo {author} {\bibfnamefont {R.}~\bibnamefont {Alicki}}, \bibinfo {author} {\bibfnamefont {M.}~\bibnamefont {Horodecki}}, \bibinfo {author} {\bibfnamefont {P.}~\bibnamefont {Horodecki}},\ and\ \bibinfo {author} {\bibfnamefont {R.}~\bibnamefont {Horodecki}},\ }\bibfield  {title} {\bibinfo {title} {On thermal stability of topological qubit in kitaev's 4d model},\ }\href {https://doi.org/10.1142/S1230161210000023} {\bibfield  {journal} {\bibinfo  {journal} {Open Systems \& Information Dynamics}\ }\textbf {\bibinfo {volume} {17}},\ \bibinfo {pages} {1} (\bibinfo {year} {2010})}\BibitemShut {NoStop}%
\bibitem [{\citenamefont {Bravyi}\ \emph {et~al.}(2010)\citenamefont {Bravyi}, \citenamefont {Poulin},\ and\ \citenamefont {Terhal}}]{bpt_bound_prl2010}%
  \BibitemOpen
  \bibfield  {author} {\bibinfo {author} {\bibfnamefont {S.}~\bibnamefont {Bravyi}}, \bibinfo {author} {\bibfnamefont {D.}~\bibnamefont {Poulin}},\ and\ \bibinfo {author} {\bibfnamefont {B.}~\bibnamefont {Terhal}},\ }\bibfield  {title} {\bibinfo {title} {Tradeoffs for reliable quantum information storage in 2d systems},\ }\href {https://doi.org/10.1103/PhysRevLett.104.050503} {\bibfield  {journal} {\bibinfo  {journal} {Phys. Rev. Lett.}\ }\textbf {\bibinfo {volume} {104}},\ \bibinfo {pages} {050503} (\bibinfo {year} {2010})}\BibitemShut {NoStop}%
\bibitem [{\citenamefont {Flammia}\ \emph {et~al.}(2017)\citenamefont {Flammia}, \citenamefont {Haah}, \citenamefont {Kastoryano},\ and\ \citenamefont {Kim}}]{Flammia_quantum2017}%
  \BibitemOpen
  \bibfield  {author} {\bibinfo {author} {\bibfnamefont {S.~T.}\ \bibnamefont {Flammia}}, \bibinfo {author} {\bibfnamefont {J.}~\bibnamefont {Haah}}, \bibinfo {author} {\bibfnamefont {M.~J.}\ \bibnamefont {Kastoryano}},\ and\ \bibinfo {author} {\bibfnamefont {I.~H.}\ \bibnamefont {Kim}},\ }\bibfield  {title} {\bibinfo {title} {Limits on the storage of quantum information in a volume of space},\ }\href {https://doi.org/10.22331/q-2017-04-25-4} {\bibfield  {journal} {\bibinfo  {journal} {{Quantum}}\ }\textbf {\bibinfo {volume} {1}},\ \bibinfo {pages} {4} (\bibinfo {year} {2017})}\BibitemShut {NoStop}%
\bibitem [{\citenamefont {Breuckmann}\ and\ \citenamefont {Eberhardt}(2021)}]{Breuckmann_prxq2021}%
  \BibitemOpen
  \bibfield  {author} {\bibinfo {author} {\bibfnamefont {N.~P.}\ \bibnamefont {Breuckmann}}\ and\ \bibinfo {author} {\bibfnamefont {J.~N.}\ \bibnamefont {Eberhardt}},\ }\bibfield  {title} {\bibinfo {title} {Quantum low-density parity-check codes},\ }\href {https://doi.org/10.1103/PRXQuantum.2.040101} {\bibfield  {journal} {\bibinfo  {journal} {PRX Quantum}\ }\textbf {\bibinfo {volume} {2}},\ \bibinfo {pages} {040101} (\bibinfo {year} {2021})}\BibitemShut {NoStop}%
\bibitem [{\citenamefont {Tremblay}\ \emph {et~al.}(2022)\citenamefont {Tremblay}, \citenamefont {Delfosse},\ and\ \citenamefont {Beverland}}]{tremblay_prl2022}%
  \BibitemOpen
  \bibfield  {author} {\bibinfo {author} {\bibfnamefont {M.~A.}\ \bibnamefont {Tremblay}}, \bibinfo {author} {\bibfnamefont {N.}~\bibnamefont {Delfosse}},\ and\ \bibinfo {author} {\bibfnamefont {M.~E.}\ \bibnamefont {Beverland}},\ }\bibfield  {title} {\bibinfo {title} {Constant-overhead quantum error correction with thin planar connectivity},\ }\href {https://doi.org/10.1103/PhysRevLett.129.050504} {\bibfield  {journal} {\bibinfo  {journal} {Phys. Rev. Lett.}\ }\textbf {\bibinfo {volume} {129}},\ \bibinfo {pages} {050504} (\bibinfo {year} {2022})}\BibitemShut {NoStop}%
\bibitem [{\citenamefont {Bravyi}\ \emph {et~al.}(2024)\citenamefont {Bravyi}, \citenamefont {Cross}, \citenamefont {Gambetta}, \citenamefont {Maslov}, \citenamefont {Rall},\ and\ \citenamefont {Yoder}}]{Bravyi_nature2024}%
  \BibitemOpen
  \bibfield  {author} {\bibinfo {author} {\bibfnamefont {S.}~\bibnamefont {Bravyi}}, \bibinfo {author} {\bibfnamefont {A.~W.}\ \bibnamefont {Cross}}, \bibinfo {author} {\bibfnamefont {J.~M.}\ \bibnamefont {Gambetta}}, \bibinfo {author} {\bibfnamefont {D.}~\bibnamefont {Maslov}}, \bibinfo {author} {\bibfnamefont {P.}~\bibnamefont {Rall}},\ and\ \bibinfo {author} {\bibfnamefont {T.~J.}\ \bibnamefont {Yoder}},\ }\bibfield  {title} {\bibinfo {title} {High-threshold and low-overhead fault-tolerant quantum memory},\ }\href {https://doi.org/10.1038/s41586-024-07107-7} {\bibfield  {journal} {\bibinfo  {journal} {Nature}\ }\textbf {\bibinfo {volume} {627}},\ \bibinfo {pages} {778} (\bibinfo {year} {2024})}\BibitemShut {NoStop}%
\bibitem [{\citenamefont {Bluvstein}\ \emph {et~al.}(2022)\citenamefont {Bluvstein}, \citenamefont {Levine}, \citenamefont {Semeghini}, \citenamefont {Wang}, \citenamefont {Ebadi}, \citenamefont {Kalinowski}, \citenamefont {Keesling}, \citenamefont {Maskara}, \citenamefont {Pichler}, \citenamefont {Greiner}, \citenamefont {Vuleti{\'{c}}},\ and\ \citenamefont {Lukin}}]{Bluvstein_nature2022}%
  \BibitemOpen
  \bibfield  {author} {\bibinfo {author} {\bibfnamefont {D.}~\bibnamefont {Bluvstein}}, \bibinfo {author} {\bibfnamefont {H.}~\bibnamefont {Levine}}, \bibinfo {author} {\bibfnamefont {G.}~\bibnamefont {Semeghini}}, \bibinfo {author} {\bibfnamefont {T.~T.}\ \bibnamefont {Wang}}, \bibinfo {author} {\bibfnamefont {S.}~\bibnamefont {Ebadi}}, \bibinfo {author} {\bibfnamefont {M.}~\bibnamefont {Kalinowski}}, \bibinfo {author} {\bibfnamefont {A.}~\bibnamefont {Keesling}}, \bibinfo {author} {\bibfnamefont {N.}~\bibnamefont {Maskara}}, \bibinfo {author} {\bibfnamefont {H.}~\bibnamefont {Pichler}}, \bibinfo {author} {\bibfnamefont {M.}~\bibnamefont {Greiner}}, \bibinfo {author} {\bibfnamefont {V.}~\bibnamefont {Vuleti{\'{c}}}},\ and\ \bibinfo {author} {\bibfnamefont {M.~D.}\ \bibnamefont {Lukin}},\ }\bibfield  {title} {\bibinfo {title} {A quantum processor based on coherent transport of entangled atom arrays},\ }\href {https://doi.org/10.1038/s41586-022-04592-6} {\bibfield  {journal} {\bibinfo  {journal} {Nature}\
  }\textbf {\bibinfo {volume} {604}},\ \bibinfo {pages} {451} (\bibinfo {year} {2022})}\BibitemShut {NoStop}%
\bibitem [{\citenamefont {Bluvstein}\ \emph {et~al.}(2024)\citenamefont {Bluvstein}, \citenamefont {Evered}, \citenamefont {Geim}, \citenamefont {Li}, \citenamefont {Zhou}, \citenamefont {Manovitz}, \citenamefont {Ebadi}, \citenamefont {Cain}, \citenamefont {Kalinowski}, \citenamefont {Hangleiter}, \citenamefont {Bonilla~Ataides}, \citenamefont {Maskara}, \citenamefont {Cong}, \citenamefont {Gao}, \citenamefont {Sales~Rodriguez}, \citenamefont {Karolyshyn}, \citenamefont {Semeghini}, \citenamefont {Gullans}, \citenamefont {Greiner}, \citenamefont {Vuleti{\'{c}}},\ and\ \citenamefont {Lukin}}]{Bluvstein_nature2024}%
  \BibitemOpen
  \bibfield  {author} {\bibinfo {author} {\bibfnamefont {D.}~\bibnamefont {Bluvstein}}, \bibinfo {author} {\bibfnamefont {S.~J.}\ \bibnamefont {Evered}}, \bibinfo {author} {\bibfnamefont {A.~A.}\ \bibnamefont {Geim}}, \bibinfo {author} {\bibfnamefont {S.~H.}\ \bibnamefont {Li}}, \bibinfo {author} {\bibfnamefont {H.}~\bibnamefont {Zhou}}, \bibinfo {author} {\bibfnamefont {T.}~\bibnamefont {Manovitz}}, \bibinfo {author} {\bibfnamefont {S.}~\bibnamefont {Ebadi}}, \bibinfo {author} {\bibfnamefont {M.}~\bibnamefont {Cain}}, \bibinfo {author} {\bibfnamefont {M.}~\bibnamefont {Kalinowski}}, \bibinfo {author} {\bibfnamefont {D.}~\bibnamefont {Hangleiter}}, \bibinfo {author} {\bibfnamefont {J.~P.}\ \bibnamefont {Bonilla~Ataides}}, \bibinfo {author} {\bibfnamefont {N.}~\bibnamefont {Maskara}}, \bibinfo {author} {\bibfnamefont {I.}~\bibnamefont {Cong}}, \bibinfo {author} {\bibfnamefont {X.}~\bibnamefont {Gao}}, \bibinfo {author} {\bibfnamefont {P.}~\bibnamefont {Sales~Rodriguez}}, \bibinfo {author} {\bibfnamefont
  {T.}~\bibnamefont {Karolyshyn}}, \bibinfo {author} {\bibfnamefont {G.}~\bibnamefont {Semeghini}}, \bibinfo {author} {\bibfnamefont {M.~J.}\ \bibnamefont {Gullans}}, \bibinfo {author} {\bibfnamefont {M.}~\bibnamefont {Greiner}}, \bibinfo {author} {\bibfnamefont {V.}~\bibnamefont {Vuleti{\'{c}}}},\ and\ \bibinfo {author} {\bibfnamefont {M.~D.}\ \bibnamefont {Lukin}},\ }\bibfield  {title} {\bibinfo {title} {Logical quantum processor based on reconfigurable atom arrays},\ }\href {https://doi.org/10.1038/s41586-023-06927-3} {\bibfield  {journal} {\bibinfo  {journal} {Nature}\ }\textbf {\bibinfo {volume} {626}},\ \bibinfo {pages} {58} (\bibinfo {year} {2024})}\BibitemShut {NoStop}%
\bibitem [{\citenamefont {Xu}\ \emph {et~al.}(2024)\citenamefont {Xu}, \citenamefont {Bonilla~Ataides}, \citenamefont {Pattison}, \citenamefont {Raveendran}, \citenamefont {Bluvstein}, \citenamefont {Wurtz}, \citenamefont {Vasi{\'{c}}}, \citenamefont {Lukin}, \citenamefont {Jiang},\ and\ \citenamefont {Zhou}}]{xu_nature2024}%
  \BibitemOpen
  \bibfield  {author} {\bibinfo {author} {\bibfnamefont {Q.}~\bibnamefont {Xu}}, \bibinfo {author} {\bibfnamefont {J.~P.}\ \bibnamefont {Bonilla~Ataides}}, \bibinfo {author} {\bibfnamefont {C.~A.}\ \bibnamefont {Pattison}}, \bibinfo {author} {\bibfnamefont {N.}~\bibnamefont {Raveendran}}, \bibinfo {author} {\bibfnamefont {D.}~\bibnamefont {Bluvstein}}, \bibinfo {author} {\bibfnamefont {J.}~\bibnamefont {Wurtz}}, \bibinfo {author} {\bibfnamefont {B.}~\bibnamefont {Vasi{\'{c}}}}, \bibinfo {author} {\bibfnamefont {M.~D.}\ \bibnamefont {Lukin}}, \bibinfo {author} {\bibfnamefont {L.}~\bibnamefont {Jiang}},\ and\ \bibinfo {author} {\bibfnamefont {H.}~\bibnamefont {Zhou}},\ }\bibfield  {title} {\bibinfo {title} {Constant-overhead fault-tolerant quantum computation with reconfigurable atom arrays},\ }\href {https://doi.org/10.1038/s41567-024-02479-z} {\bibfield  {journal} {\bibinfo  {journal} {Nature Physics}\ }\textbf {\bibinfo {volume} {20}},\ \bibinfo {pages} {1084} (\bibinfo {year} {2024})}\BibitemShut {NoStop}%
\bibitem [{\citenamefont {Krastanov}\ \emph {et~al.}(2025)\citenamefont {Krastanov}, \citenamefont {Mian}, \citenamefont {Viswanathan}, \citenamefont {Pardis}, \citenamefont {Lapeyre}, \citenamefont {Micciche}, \citenamefont {Yan}, \citenamefont {gsommers}, \citenamefont {Hofmann}, \citenamefont {Preiß}, \citenamefont {Bhatt}, \citenamefont {IsaacP1234}, \citenamefont {Göttgens}, \citenamefont {Meligrana}, \citenamefont {Benzillaist}, \citenamefont {Zhao}, \citenamefont {ShuGe-MIT}, \citenamefont {Holy}, \citenamefont {Dang}, \citenamefont {adrianariton},\ and\ \citenamefont {ismoldayev}}]{QuantumClifford}%
  \BibitemOpen
  \bibfield  {author} {\bibinfo {author} {\bibfnamefont {S.}~\bibnamefont {Krastanov}}, \bibinfo {author} {\bibfnamefont {F.~A.}\ \bibnamefont {Mian}}, \bibinfo {author} {\bibfnamefont {P.}~\bibnamefont {Viswanathan}}, \bibinfo {author} {\bibfnamefont {S.}~\bibnamefont {Pardis}}, \bibinfo {author} {\bibfnamefont {J.}~\bibnamefont {Lapeyre}}, \bibinfo {author} {\bibfnamefont {A.}~\bibnamefont {Micciche}}, \bibinfo {author} {\bibfnamefont {Y.}~\bibnamefont {Yan}}, \bibinfo {author} {\bibnamefont {gsommers}}, \bibinfo {author} {\bibfnamefont {T.}~\bibnamefont {Hofmann}}, \bibinfo {author} {\bibfnamefont {Q.}~\bibnamefont {Preiß}}, \bibinfo {author} {\bibfnamefont {A.}~\bibnamefont {Bhatt}}, \bibinfo {author} {\bibnamefont {IsaacP1234}}, \bibinfo {author} {\bibfnamefont {L.}~\bibnamefont {Göttgens}}, \bibinfo {author} {\bibfnamefont {A.}~\bibnamefont {Meligrana}}, \bibinfo {author} {\bibnamefont {Benzillaist}}, \bibinfo {author} {\bibfnamefont {C.}~\bibnamefont {Zhao}}, \bibinfo {author} {\bibnamefont
  {ShuGe-MIT}}, \bibinfo {author} {\bibfnamefont {T.}~\bibnamefont {Holy}}, \bibinfo {author} {\bibfnamefont {T.}~\bibnamefont {Dang}}, \bibinfo {author} {\bibnamefont {adrianariton}},\ and\ \bibinfo {author} {\bibnamefont {ismoldayev}},\ }\href {https://doi.org/10.5281/zenodo.15178535} {\bibinfo {title} {Quantumsavory/quantumclifford.jl: v0.9.19}} (\bibinfo {year} {2025})\BibitemShut {NoStop}%
\bibitem [{\citenamefont {Nahum}\ \emph {et~al.}(2018)\citenamefont {Nahum}, \citenamefont {Vijay},\ and\ \citenamefont {Haah}}]{Nahum2017op}%
  \BibitemOpen
  \bibfield  {author} {\bibinfo {author} {\bibfnamefont {A.}~\bibnamefont {Nahum}}, \bibinfo {author} {\bibfnamefont {S.}~\bibnamefont {Vijay}},\ and\ \bibinfo {author} {\bibfnamefont {J.}~\bibnamefont {Haah}},\ }\bibfield  {title} {\bibinfo {title} {{Operator Spreading in Random Unitary Circuits}},\ }\href {https://doi.org/10.1103/PhysRevX.8.021014} {\bibfield  {journal} {\bibinfo  {journal} {Phys. Rev. X}\ }\textbf {\bibinfo {volume} {8}},\ \bibinfo {pages} {021014} (\bibinfo {year} {2018})}\BibitemShut {NoStop}%
\bibitem [{\citenamefont {Bertini}\ \emph {et~al.}(2019)\citenamefont {Bertini}, \citenamefont {Kos},\ and\ \citenamefont {Prosen}}]{Bertini2019}%
  \BibitemOpen
  \bibfield  {author} {\bibinfo {author} {\bibfnamefont {B.}~\bibnamefont {Bertini}}, \bibinfo {author} {\bibfnamefont {P.}~\bibnamefont {Kos}},\ and\ \bibinfo {author} {\bibfnamefont {T.}~\bibnamefont {Prosen}},\ }\bibfield  {title} {\bibinfo {title} {{Exact Correlation Functions for Dual-Unitary Lattice Models in $1+1$ Dimensions}},\ }\href {https://doi.org/10.1103/PhysRevLett.123.210601} {\bibfield  {journal} {\bibinfo  {journal} {Physical Review Letters}\ }\textbf {\bibinfo {volume} {123}},\ \bibinfo {pages} {210601} (\bibinfo {year} {2019})},\ \Eprint {https://arxiv.org/abs/1904.02140} {1904.02140} \BibitemShut {NoStop}%
\bibitem [{\citenamefont {Piroli}\ \emph {et~al.}(2020)\citenamefont {Piroli}, \citenamefont {Bertini}, \citenamefont {Cirac},\ and\ \citenamefont {Prosen}}]{Piroli2019}%
  \BibitemOpen
  \bibfield  {author} {\bibinfo {author} {\bibfnamefont {L.}~\bibnamefont {Piroli}}, \bibinfo {author} {\bibfnamefont {B.}~\bibnamefont {Bertini}}, \bibinfo {author} {\bibfnamefont {J.~I.}\ \bibnamefont {Cirac}},\ and\ \bibinfo {author} {\bibfnamefont {T.~c.~v.}\ \bibnamefont {Prosen}},\ }\bibfield  {title} {\bibinfo {title} {Exact dynamics in dual-unitary quantum circuits},\ }\href {https://doi.org/10.1103/PhysRevB.101.094304} {\bibfield  {journal} {\bibinfo  {journal} {Phys. Rev. B}\ }\textbf {\bibinfo {volume} {101}},\ \bibinfo {pages} {094304} (\bibinfo {year} {2020})}\BibitemShut {NoStop}%
\bibitem [{\citenamefont {Fisher}(1984)}]{Fisher_1984}%
  \BibitemOpen
  \bibfield  {author} {\bibinfo {author} {\bibfnamefont {M.~E.}\ \bibnamefont {Fisher}},\ }\bibfield  {title} {\bibinfo {title} {Walks, walls, wetting, and melting},\ }\href {https://doi.org/10.1007/BF01009436} {\bibfield  {journal} {\bibinfo  {journal} {Journal of Statistical Physics}\ }\textbf {\bibinfo {volume} {34}},\ \bibinfo {pages} {667} (\bibinfo {year} {1984})}\BibitemShut {NoStop}%
\bibitem [{\citenamefont {Fisher}(1986)}]{Fisher_5-epsilon_1986}%
  \BibitemOpen
  \bibfield  {author} {\bibinfo {author} {\bibfnamefont {D.~S.}\ \bibnamefont {Fisher}},\ }\bibfield  {title} {\bibinfo {title} {Interface fluctuations in disordered systems: $5\ensuremath{-}\ensuremath{\epsilon}$ expansion and failure of dimensional reduction},\ }\href {https://doi.org/10.1103/PhysRevLett.56.1964} {\bibfield  {journal} {\bibinfo  {journal} {Phys. Rev. Lett.}\ }\textbf {\bibinfo {volume} {56}},\ \bibinfo {pages} {1964} (\bibinfo {year} {1986})}\BibitemShut {NoStop}%
\bibitem [{\citenamefont {Dembin}\ \emph {et~al.}(2025{\natexlab{a}})\citenamefont {Dembin}, \citenamefont {Elboim}, \citenamefont {Hadas},\ and\ \citenamefont {Peled}}]{Peled2401}%
  \BibitemOpen
  \bibfield  {author} {\bibinfo {author} {\bibfnamefont {B.}~\bibnamefont {Dembin}}, \bibinfo {author} {\bibfnamefont {D.}~\bibnamefont {Elboim}}, \bibinfo {author} {\bibfnamefont {D.}~\bibnamefont {Hadas}},\ and\ \bibinfo {author} {\bibfnamefont {R.}~\bibnamefont {Peled}},\ }\href {https://arxiv.org/abs/2401.06768} {\bibinfo {title} {Minimal surfaces in random environment}} (\bibinfo {year} {2025}{\natexlab{a}}),\ \Eprint {https://arxiv.org/abs/2401.06768} {arXiv:2401.06768 [math-ph]} \BibitemShut {NoStop}%
\bibitem [{\citenamefont {Aizenman}\ and\ \citenamefont {Wehr}(1989)}]{Aizenman_Wehr_prl1989_ImryMa}%
  \BibitemOpen
  \bibfield  {author} {\bibinfo {author} {\bibfnamefont {M.}~\bibnamefont {Aizenman}}\ and\ \bibinfo {author} {\bibfnamefont {J.}~\bibnamefont {Wehr}},\ }\bibfield  {title} {\bibinfo {title} {Rounding of first-order phase transitions in systems with quenched disorder},\ }\href {https://doi.org/10.1103/PhysRevLett.62.2503} {\bibfield  {journal} {\bibinfo  {journal} {Phys. Rev. Lett.}\ }\textbf {\bibinfo {volume} {62}},\ \bibinfo {pages} {2503} (\bibinfo {year} {1989})}\BibitemShut {NoStop}%
\bibitem [{\citenamefont {Aizenman}\ and\ \citenamefont {Wehr}(1990)}]{Aizenman_Wehr_ImryMa}%
  \BibitemOpen
  \bibfield  {author} {\bibinfo {author} {\bibfnamefont {M.}~\bibnamefont {Aizenman}}\ and\ \bibinfo {author} {\bibfnamefont {J.}~\bibnamefont {Wehr}},\ }\bibfield  {title} {\bibinfo {title} {Rounding effects of quenched randomness on first-order phase transitions},\ }\href {https://doi.org/10.1007/BF02096933} {\bibfield  {journal} {\bibinfo  {journal} {Communications in Mathematical Physics}\ }\textbf {\bibinfo {volume} {130}},\ \bibinfo {pages} {489} (\bibinfo {year} {1990})}\BibitemShut {NoStop}%
\bibitem [{\citenamefont {Wiese}(2022)}]{Wiese2022}%
  \BibitemOpen
  \bibfield  {author} {\bibinfo {author} {\bibfnamefont {K.~J.}\ \bibnamefont {Wiese}},\ }\bibfield  {title} {\bibinfo {title} {{Theory and experiments for disordered elastic manifolds, depinning, avalanches, and sandpiles}},\ }\href {https://doi.org/10.1088/1361-6633/ac4648} {\bibfield  {journal} {\bibinfo  {journal} {Reports on Progress in Physics}\ }\textbf {\bibinfo {volume} {85}},\ \bibinfo {pages} {086502} (\bibinfo {year} {2022})},\ \Eprint {https://arxiv.org/abs/2102.01215} {2102.01215} \BibitemShut {NoStop}%
\bibitem [{\citenamefont {Alava}\ and\ \citenamefont {Duxbury}(1996)}]{Alava_Duxbury_2+1rough1996}%
  \BibitemOpen
  \bibfield  {author} {\bibinfo {author} {\bibfnamefont {M.~J.}\ \bibnamefont {Alava}}\ and\ \bibinfo {author} {\bibfnamefont {P.~M.}\ \bibnamefont {Duxbury}},\ }\bibfield  {title} {\bibinfo {title} {Disorder-induced roughening in the three-dimensional ising model},\ }\href {https://doi.org/10.1103/PhysRevB.54.14990} {\bibfield  {journal} {\bibinfo  {journal} {Phys. Rev. B}\ }\textbf {\bibinfo {volume} {54}},\ \bibinfo {pages} {14990} (\bibinfo {year} {1996})}\BibitemShut {NoStop}%
\bibitem [{\citenamefont {Sepp\"al\"a}\ \emph {et~al.}(2001)\citenamefont {Sepp\"al\"a}, \citenamefont {Alava},\ and\ \citenamefont {Duxbury}}]{Seppala_Duxbury_2+1rougheningtransition2001}%
  \BibitemOpen
  \bibfield  {author} {\bibinfo {author} {\bibfnamefont {E.~T.}\ \bibnamefont {Sepp\"al\"a}}, \bibinfo {author} {\bibfnamefont {M.~J.}\ \bibnamefont {Alava}},\ and\ \bibinfo {author} {\bibfnamefont {P.~M.}\ \bibnamefont {Duxbury}},\ }\bibfield  {title} {\bibinfo {title} {Intermittence and roughening of periodic elastic media},\ }\href {https://doi.org/10.1103/PhysRevE.63.036126} {\bibfield  {journal} {\bibinfo  {journal} {Phys. Rev. E}\ }\textbf {\bibinfo {volume} {63}},\ \bibinfo {pages} {036126} (\bibinfo {year} {2001})}\BibitemShut {NoStop}%
\bibitem [{\citenamefont {van~de Wetering}(2020)}]{VandeWetering2020}%
  \BibitemOpen
  \bibfield  {author} {\bibinfo {author} {\bibfnamefont {J.}~\bibnamefont {van~de Wetering}},\ }\href {http://arxiv.org/abs/2012.13966} {\bibinfo {title} {{ZX-calculus for the working quantum computer scientist}}} (\bibinfo {year} {2020}),\ \Eprint {https://arxiv.org/abs/2012.13966} {arXiv:2012.13966} \BibitemShut {NoStop}%
\bibitem [{\citenamefont {Dembin}\ \emph {et~al.}(2025{\natexlab{b}})\citenamefont {Dembin}, \citenamefont {Elboim},\ and\ \citenamefont {Peled}}]{Peled2504}%
  \BibitemOpen
  \bibfield  {author} {\bibinfo {author} {\bibfnamefont {B.}~\bibnamefont {Dembin}}, \bibinfo {author} {\bibfnamefont {D.}~\bibnamefont {Elboim}},\ and\ \bibinfo {author} {\bibfnamefont {R.}~\bibnamefont {Peled}},\ }\href {https://arxiv.org/abs/2504.10379} {\bibinfo {title} {Minimal surfaces in strongly correlated random environments}} (\bibinfo {year} {2025}{\natexlab{b}}),\ \Eprint {https://arxiv.org/abs/2504.10379} {arXiv:2504.10379 [math.PR]} \BibitemShut {NoStop}%
\bibitem [{\citenamefont {Imry}\ and\ \citenamefont {Ma}(1975)}]{Imry_Ma_prl1975}%
  \BibitemOpen
  \bibfield  {author} {\bibinfo {author} {\bibfnamefont {Y.}~\bibnamefont {Imry}}\ and\ \bibinfo {author} {\bibfnamefont {S.-k.}\ \bibnamefont {Ma}},\ }\bibfield  {title} {\bibinfo {title} {Random-field instability of the ordered state of continuous symmetry},\ }\href {https://doi.org/10.1103/PhysRevLett.35.1399} {\bibfield  {journal} {\bibinfo  {journal} {Phys. Rev. Lett.}\ }\textbf {\bibinfo {volume} {35}},\ \bibinfo {pages} {1399} (\bibinfo {year} {1975})}\BibitemShut {NoStop}%
\bibitem [{\citenamefont {Grinstein}\ and\ \citenamefont {Ma}(1983)}]{grinstein_ma_1983}%
  \BibitemOpen
  \bibfield  {author} {\bibinfo {author} {\bibfnamefont {G.}~\bibnamefont {Grinstein}}\ and\ \bibinfo {author} {\bibfnamefont {S.-k.}\ \bibnamefont {Ma}},\ }\bibfield  {title} {\bibinfo {title} {Surface tension, roughening, and lower critical dimension in the random-field ising model},\ }\href {https://doi.org/10.1103/PhysRevB.28.2588} {\bibfield  {journal} {\bibinfo  {journal} {Phys. Rev. B}\ }\textbf {\bibinfo {volume} {28}},\ \bibinfo {pages} {2588} (\bibinfo {year} {1983})}\BibitemShut {NoStop}%
\bibitem [{\citenamefont {Villain}(1984)}]{villain_prl1984}%
  \BibitemOpen
  \bibfield  {author} {\bibinfo {author} {\bibfnamefont {J.}~\bibnamefont {Villain}},\ }\bibfield  {title} {\bibinfo {title} {Nonequilibrium "critical" exponents in the random-field ising model},\ }\href {https://doi.org/10.1103/PhysRevLett.52.1543} {\bibfield  {journal} {\bibinfo  {journal} {Phys. Rev. Lett.}\ }\textbf {\bibinfo {volume} {52}},\ \bibinfo {pages} {1543} (\bibinfo {year} {1984})}\BibitemShut {NoStop}%
\bibitem [{\citenamefont {Gopalakrishnan}\ and\ \citenamefont {Lamacraft}(2019)}]{Gopalakrishnan2019}%
  \BibitemOpen
  \bibfield  {author} {\bibinfo {author} {\bibfnamefont {S.}~\bibnamefont {Gopalakrishnan}}\ and\ \bibinfo {author} {\bibfnamefont {A.}~\bibnamefont {Lamacraft}},\ }\bibfield  {title} {\bibinfo {title} {{Unitary circuits of finite depth and infinite width from quantum channels}},\ }\href {https://doi.org/10.1103/PhysRevB.100.064309} {\bibfield  {journal} {\bibinfo  {journal} {Physical Review B}\ }\textbf {\bibinfo {volume} {100}},\ \bibinfo {pages} {064309} (\bibinfo {year} {2019})},\ \Eprint {https://arxiv.org/abs/1903.11611} {1903.11611} \BibitemShut {NoStop}%
\bibitem [{Note4()}]{Note4}%
  \BibitemOpen
  \bibinfo {note} {We stress that $t$ and $x$ are just the numbers with a unit of lattice constant.}\BibitemShut {Stop}%
\end{thebibliography}%


\newpage
\onecolumngrid

\begin{center}
    \textbf{\large Supplemental Material for \\
``Roughening Transition in Quantum Circuits''}
    
    \vspace{0.5cm}

    \author{Hyunsoo Ha}
    \author{David A. Huse}
    \author{Grace M. Sommers}
    \affiliation{Department of Physics, Princeton University, Princeton, NJ 08544, USA}

    Hyunsoo Ha$^1$, David A. Huse$^1$, Grace M. Sommers$^{1}$
    
    \vspace{0.2cm}
    
    $^1$\textit{Department of Physics, Princeton University, Princeton, New Jersey 08544, USA} 
\end{center}

\vspace{1cm}

\setcounter{section}{0}
\setcounter{figure}{0}
\setcounter{equation}{0}
\setcounter{page}{1}
\makeatletter
\renewcommand{\thefigure}{S\arabic{figure}}
\renewcommand{\theequation}{S\arabic{equation}}
\renewcommand{\thesection}{S\arabic{section}}
\makeatother
\setcounter{secnumdepth}{2}

\begin{center}
\begin{minipage}{0.85\textwidth}
\vspace{-1 cm}
In this Supplemental Material, we provide (I) a review of the entanglement membrane in entanglement growth and known results on membrane roughening from classical statistical mechanics in various dimensions, (II) an explanation of the hyperdiamond circuit architecture using the ZX-calculus, (III) a detailed discussion of the scaling behavior of entanglement entropy under tilts away from the membrane-pinning orientation, and (IV) an analysis of the fluctuations of entanglement entropy under tilt across different disorder realizations.
\vspace{0.5cm}
\end{minipage}
\end{center}

\section{Preliminaries}
\subsection{Entanglement Membrane}
Entanglement membrane theory is an effective statistical mechanics model of entanglement growth in local quantum systems, which posits that the entropy of a subsystem is given by the free energy of a spacetime interface separating the subsystem from its complement~\cite{Nahum2017op,Nahum_Haah_entanglementgrowth_2016,Jonay2018,Zhou2019, Zhou2020,sierant_turkeshi_membrane2023}. When the dynamics generates volume law entanglement, the interface has a nonzero tension $\mathcal{E}(\hat{\mathbf{n}})$, where $\mathbf{\hat{n}}$ is the vector normal to the interface.

While the microscopic derivation of the theory typically relies on an \textit{ensemble} of random circuits, it can also be extended to \textit{individual} realizations of the dynamics~\cite{Zhou2020,Sommers_Huse_2024}. In particular, the tension has a geometrical interpretation as the density of information flow across the membrane, and thus depends on the nature of the dynamics in the direction $\mathbf{\hat{n}}$. When this dynamics is unitary (unitary within a subspace), the membrane free energy is simply the dimension of the full Hilbert space (dimension of the subspace). \textit{Fully multi-unitary} models are those for which every possible membrane orientation is spacelike with respect to some unitary arrow of time. In such cases, the free energy is a piecewise-linear function of the membrane orientation, and in that sense the membrane behaves like an interface in a zero-temperature lattice model.

In a random Clifford circuit (composed of unitary Clifford gates and/or stabilizer projectors), the dynamics across an arbitrarily oriented membrane is not strictly unitary, but as argued in Ref.~\cite{Sommers_Huse_2024}, the membrane is still described by a zero-temperature lattice model, albeit one with disorder. Consider the fate of a fully mixed initial state evolved in the direction $\hat{\mathbf{n}}$. If the dynamics in that direction is in a volume-law phase, the entropy density reaches an \textit{emergent plateau}  with some nonzero entropy density $s$ on a time scale that is polynomial in the system size.  This is followed by occasional purification events (whose rate becomes exponentially small in the system size) which reduce the entropy by one bit extra below the ``plateau''.  But these purification events are discrete and rare, corresponding in the entanglement membrane picture to the membrane finding a lower-energy place to sit as it is given more room to wander. In contrast, in a finite-temperature lattice model, the free energy of an interface includes contributions from many different paths, not just the ones with the lowest energy.

Let us briefly review the situation in (1+1)d. The statistical mechanics model for membranes of random hybrid Clifford circuits in the mixed/volume-law phase~\cite{Li_Vijay_Fisher_DPRE2023}, as well as timelike membranes in random Clifford circuits without measurements~\cite{Nahum_Haah_entanglementgrowth_2016,Zhou2019}, is the directed polymer in a random environment (DPRE) at zero temperature.  For certain space-time-translationally-invariant (STTI) circuits, the membrane may be pinned to the lattice and remain smooth.  But for any nonzero amount of disorder, these membranes are rough: a membrane of length $l$ has transverse displacements away from smooth $\sim l^{\zeta}$ with $\zeta = 2/3$, and the free energy (the entanglement entropy for the circuit models) has a subleading contribution $\sim l^{\theta}$ with $\theta=1/3$. It is worth noting that temperature is an irrelevant perturbation to the DPRE fixed point~\cite{Halpin-Healy_FRG_1990}, so the microscopically nonzero-temperature DPRE describing entanglement membranes in (1+1)d Haar-random circuits is described by the same zero-temperature fixed point in the thermodynamic limit.

Dual-unitary circuits are a special class of (1+1)d circuits which fall under the broader umbrella of fully multi-unitary models discussed above. These circuits are unitary in two orthogonal directions (the standard time direction, and the space direction)~\cite{Gopalakrishnan2019,Bertini2019,Piroli2019}, which imposes~\cite{Zhou2020}
\begin{equation}
S(x,t) = \max(|x|, t)~,
\end{equation}
which means the line tension is $\mathcal{E}(\hat{\mathbf{n}})= \max(|n_x|,|n_t|)$.  This form of the line tension holds even if the circuit contains disorder, as long as dual-unitarity is preserved, i.e., each two-qubit gate is drawn from the ensemble of dual-unitary gates. The same concept applies in higher dimensions, so when introducing disorder into our (3+1)d circuits, we choose the disorder to destroy multi-unitarity in order to capture more generic behavior.

\subsection{Roughening induced by disorder}
This section reviews possible fixed points of a membrane at zero temperature, both with and without disorder and/or periodic lattice potentials, in various dimensions.  We specifically consider a $d$-dimensional membrane in a $(d+1)$-dimensional space (or spacetime in the case of our circuits), represented by heights $h(\vec{r})$ defined over a $d$-dimensional spatial domain ($\vec{r} \in \mathbb{R}^d$). This membrane can be an interface in the ordered phase of an Ising model, subject to specific boundary conditions~\cite{Fisher_1984}: the boundary spins are fixed to point up in the top half and down in the bottom half.  This setup creates a domain wall that separates two regions with opposite spin orientations. Disorder can be introduced through a random potential $V_d(\vec{r}, h)$, which is independently and identically distributed (i.i.d.) across spatial locations and heights (across spacetime locations for our circuits). This corresponds to randomness in the bonds of the Ising model. A periodic lattice potential can also be included in the form $V_p(h) = V_p(h + a)$, or simply adding $V_p\sim\cos(2\pi h/a)$, where $a$ is the lattice constant along the height direction.

For a finite system of length scale $L$ (membrane area/volume $L^d$), the membrane is smooth if the deviations $\delta h$ of the height from a perfectly flat membrane remain finite at large $L$: $\delta h \sim O(L^0)$.  The membrane is rough if the standard deviation of the height field $\delta h$ diverges with increasing $L$.  This can be a power law divergence $\sim L^\zeta$ with a positive wandering exponent $\zeta>0$, or a weaker logarithmic divergence.

A trivial flat (constant $h$) fixed point (1) occurs in the absence of both disorder and lattice potentials in the continuum. This is a zero-temperature smooth nonrandom fixed point, which we take as a reference. In this case, the continuous translational symmetry along the height direction is broken, giving rise to a corresponding ``Goldstone mode''. One way of seeing this continuous symmetry is by imposing tilted boundary conditions that impose a weak tilt $h(\vec{r})=\vec{\tau}\cdot\vec{r}$ with $|\vec{\tau}|\ll 1$ on the membrane at its boundary and seeing that this same tilt occurs for the membrane in the bulk.

Next, we consider the case where lattice potentials are present, but there is still no randomness. This perturbation is relevant at the previously discussed fixed point and reduces the continuous translational symmetry along the height direction to discrete, allowing only integer shifts in height (at the minimum of the periodic potential). This is a zero-temperature, smooth, nonrandom pinned fixed point (2). It is distinct from fixed point (1) as it is \textit{gapped}, meaning it does not support a Goldstone mode and responds to tilted boundary conditions by having discrete steps instead of uniformly tilting in the bulk.  Such behavior is present for the entanglement membrane in certain nonrandom space-time-translational-invariant (STTI) circuits with a volume-law entangled phase~\cite{Sommers_Huse_2024, Sommers_Huse_Gullans_2023}.

Returning to the first fixed point, we now introduce a random potential, which is also a relevant perturbation. For $(4+1)$ dimensions and lower \cite{Huse_Henley_1985,Huse_Henley_Fisher_respond_1985,Fisher_5-epsilon_1986,Halpin-Healy_FRG_1990,Peled2401}, by an adaptation of an Imry-Ma argument (comparing $d-2$ to $d/2$)~\cite{Aizenman_Wehr_prl1989_ImryMa, Aizenman_Wehr_ImryMa,Wiese2022}; this disorder roughens the interface, leading to a zero-temperature rough randomly-pinned fixed point (3). Since the interface is rough, it nonlinearly probes the random potential along the height direction, making this fixed point non-Gaussian. In $(4+1)$ dimensions, the roughness is only logarithmic but still non-Gaussian. For dimensions higher than $(4+1)$d, the interface remains smooth for weak nonzero disorder. However, the randomness breaks the continuous symmetry, distinguishing this fixed point from the trivial smooth fixed point. In this case, the system reaches a zero-temperature smooth randomly-pinned Gaussian critical fixed point (4).

So far, we have considered the effects of either a lattice potential or disorder separately, each leading to distinct fixed points. Now, we examine cases where both ingredients are present simultaneously by introducing a random potential perturbatively to the smooth, nonrandom pinned fixed point (2). For dimensions lower than $(2+1)$d, the random potential is relevant due to an adaptation of the Imry-Ma argument applied to a discrete symmetry (as the lattice potential discretizes the height values). 
Consequently, disorder roughens the interface and drives the system toward the rough, randomly pinned fixed point (3). In this regime, the random potential dominates over the lattice potential, and the rough fixed point remains linearly stable against the addition of a weak lattice potential. The case of $(2+1)$d is marginal~\cite{Bovier-Külske_2d}. It is generally believed that the interface remains rough in the presence of nonzero disorder, as in lower dimensions. 
However, some possibility of a transition is reported \cite{Alava_Duxbury_2+1rough1996,Seppala_Duxbury_2+1rougheningtransition2001}, suggesting that the disorder type may play a role in determining the behavior at the marginal dimension.
These rough, randomly pinned fixed points characterize the rough phase observed in random circuits for $(3+1)$ dimensions or lower \cite{Nahum_Haah_entanglementgrowth_2016, Li_Vijay_Fisher_DPRE2023, sierant_turkeshi_membrane2023}.

For dimensions higher than $(2+1)$d, adding a weak random potential at the smooth, nonrandom pinned fixed point (2) does not roughen the membrane. However, it does break the discrete translational symmetry of the system. As a result, this leads to a \textit{smooth randomly pinned gapped} fixed point (5).
In particular, for $(3+1)$d, a sufficiently strong disorder can drive the membrane to become rough \cite{Emig_Nattermann_1998_FRGprl, Emig_Nattermann_1998_FRGlong, Noh_Rieger_2002_numerics}. This roughening transition is the focus of the present work. The existence of the smooth phase in $(3+1)$d has been mathematically proven \cite{Bovier-Külske_3d,peled_2023_math_randomsurface,Peled_mathproof2023}, though a proof of the roughening transition itself has not been established. 

Finally, if we introduce a lattice potential to the smooth, randomly pinned fixed point (4), this perturbation is relevant and gaps the system, placing it within the same family of smooth, randomly pinned fixed points as (5).

\section{Hyperdiamond Circuit}

\subsection{Implementation of the generalized ($d$+1)-dimensional circuit}

In the main text, we introduced a circuit-level implementation of a hyperdiamond architecture in (3+1)-dimensions. For completeness, we present its direct generalization to arbitrary ($d$+1)-dimensions, obtained by formally replacing $3 \to d$. In this construction, $d = 1$ corresponds to the two-dimensional honeycomb lattice, and $d = 2$ corresponds to the three-dimensional diamond lattice.

For a general spatial dimension $d$, we consider qubits placed on $\textbf{r} \in \mathbb{Z}^d$. For each lattice site $\textbf{r}$, we define a set of target sites:
\begin{align}
    \mathcal{S}_{T}(\textbf{r}) = \{\textbf{r}+\textbf{e}_i|i\in\{1,\cdots,d\}\},
\end{align}
where $\textbf{e}_i$ are the unit vectors along the coordinate directions. Depending on the boundary condition, directions that leave the system may be omitted (open) or interpreted modulo the system size (periodic).

The circuit consists of single-qubit gates $G_1(t,\textbf{r})$ and two-qubit gates $G_2(t,\textbf{r}_C,\textbf{r}_T)$, which are independently and randomly selected at each realization according to the following rules:
\begin{equation}
G_1(t, \textbf{r}) = 
\begin{cases}
\hat{H}, & \text{with probability } p, \\
\hat{\mathbb{I}}, & \text{with probability } 1 - p,
\end{cases}
\end{equation}
where $\hat{H}$ is the Hadamard gate, and
\begin{equation}
G_2(t, \textbf{r}_C, \textbf{r}_T) = 
\begin{cases}
\mathrm{CNOT}(\textbf{r}_C, \textbf{r}_T), & \text{with probability } 1 - p, \\
\text{projective measurement of } \hat{Z}(\textbf{r}_C)\hat{X}(\textbf{r}_T), & \text{with probability } p.
\end{cases}
\end{equation}

As in the main text, two-qubit gates are applied at integer time steps $t \in \mathbb{Z}$. All two-qubit gates defined at a given time will commute by construction (this is transparent in the ZX-calculus formalism discussed in the next subsection), so they can be applied simultaneously. After applying the two-qubit gates at time $t$, the single-qubit gates $G_1(t,\textbf{r})$ are applied between times $t$ and $t+1$ on each site $\textbf{r}$.

Thus, once the gates are fixed for a given circuit realization, the full time-evolution operator up to time $T$ is:
\begin{align}
    U(T,p) \equiv 
    \mathcal{T}\Big[\prod_{t=1}^T
    \Big(
    \prod_{\textbf{r}}G_1(t,\textbf{r})
    \prod_{\substack{\textbf{r}\equiv t(\text{mod}2) \\ \textbf{r}' \in \mathcal{S}_{T}(\textbf{r})}} 
    G_2(t,\textbf{r},\textbf{r}')
    \Big) \Big]
\end{align}
where we alternate the two-qubit gates on odd and even control sites, so that one full period consists of two time steps, $T_p = 2$. The parity of a site $\textbf{r} = (x_1, \ldots, x_d)$ is defined by the parity of $x_1 + \cdots + x_D$, and $\mathcal{T}[\cdot]$ denotes time-ordering of the gates.
In the disorder-free limit $p=0$, the circuit reduces to a fully deterministic unitary evolution: 
\begin{align}\label{eq:disorder_free}
    U(T_p,0) =
    \prod_{\substack{\text{even}~\textbf{r} \\ \textbf{r}' \in \mathcal{S}_{T}(\textbf{r})}} 
    \text{CNOT}(\textbf{r},\textbf{r}')
    \prod_{\substack{\text{odd}~\textbf{r} \\ \textbf{r}' \in \mathcal{S}_{T}(\textbf{r})}} 
    \text{CNOT}(\textbf{r},\textbf{r}')
\end{align}
which defines an STTI multi-unitary circuit. As we discuss below, this circuit can also be viewed as defined on a $(d+1)$-dimensional hyperdiamond lattice with the full point group symmetry of that lattice.

\subsection{ZX calculus preliminaries}
To facilitate the construction and analysis of higher-dimensional circuits with crystalline symmetries, we represent them as ``ZX diagrams,'' a universal tensor network formulation of quantum circuits endowed with graphical rewrite rules. The interested reader is referred to Ref.~\cite{VandeWetering2020} for a thorough introduction to ZX calculus. Here, we introduce only the most basic ideas needed to construct our circuits.
All operations in our constructions are time-ordered, applied from bottom (early times) to top (later times) in the diagram.

ZX diagrams are composed of Z and X spiders, conventionally colored green and red, respectively:
\begin{equation}
\raisebox{-0.43\height}{\includegraphics[width=0.92\textwidth]{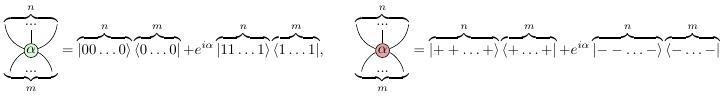}}
\label{eq:spiders}
\end{equation}
Spiders marked with no phase have $\alpha=0$. The Clifford fragment of the ZX calculus consists of diagrams where the phases on all spiders are integer multiples of $\pi/2$.

The Hadamard gate, which can be expressed non-uniquely in terms of $\pi/2$ spiders, is given its own symbol (a yellow box) for convenience:
\begin{equation}
\raisebox{-0.5\height}{\includegraphics[width=0.21\textwidth]{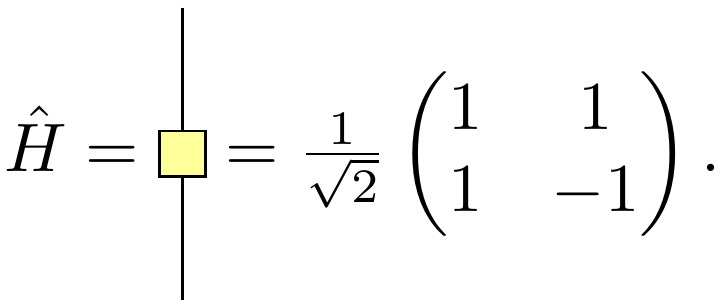}}
\label{eq:spiders}
\end{equation}

Time flows upward in our diagrams, but since the Hadamard gate is symmetric (and spiders are invariant under any permutation of legs that preserves $n$ and $m$), wires can be bent and stretched without changing the meaning of the diagram, as long as the inputs and outputs are preserved. Thus, for example, a Hadamard box can be placed along a spatial leg, as in the definition of the CZ gate:
\begin{equation}\label{eq:cz}
\raisebox{-0.5\height}{\includegraphics[width=0.18\textwidth]{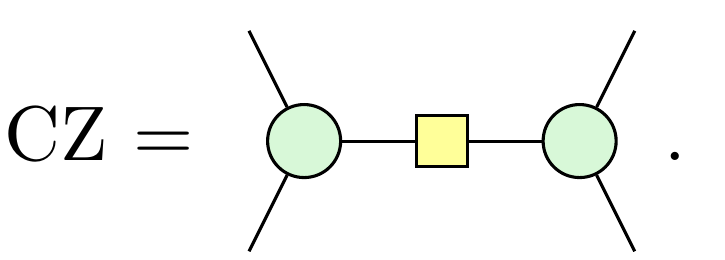}}
\end{equation}
This idea---``only connectivity matters''---is a core tenet of ZX calculus.

Two basic rules that we will use are the ``spider fusion rule'', which allows us to join spiders of the same color:
\begin{equation}\label{eq:fusion}
\raisebox{-0.5\height}{\includegraphics[width=0.21\textwidth]{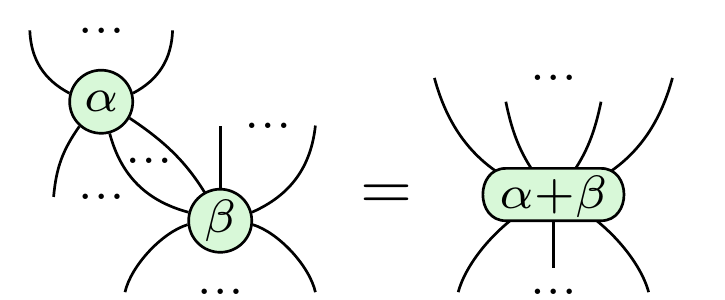}}
\end{equation}
and the 
``color-change rule,'' which allow us to push Hadamards through spiders at the cost of changing their color:
\begin{equation}\label{eq:hadamard-rule}
\raisebox{-0.5\height}{\includegraphics[width=0.21\textwidth]{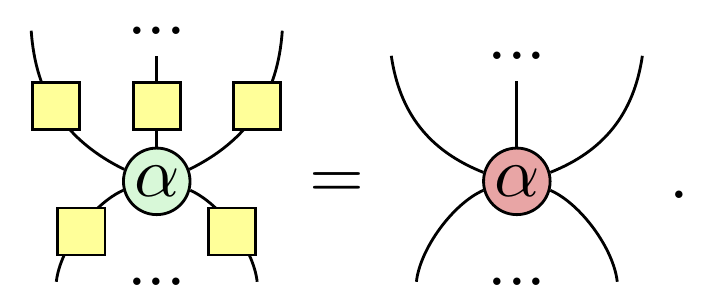}}
\end{equation}
The latter rule, combined with Eqn.~\ref{eq:cz}, gives us a simple form for the CNOT gate:
\begin{equation}\label{eq:cnot}
\raisebox{-0.5\height}{\includegraphics[width=0.21\textwidth]{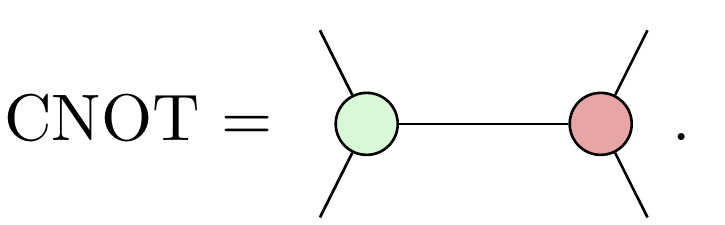}}
\end{equation}
Many-legged generalizations of CNOT are the building blocks of the isotropic hyperdiamond circuit defined below.

Before we proceed with the details of the hyperdiamond architecture, it is important to clarify the ordering of operations in the circuit. Each gate in the diagram has two types of legs: \textit{time-like legs}, which run vertically, and \textit{space-like legs}, which connect across sites at the same time. While the time-like legs clearly indicate a natural order---operations are applied from bottom to top---the space-like legs may seem ambiguous at first, since it is not obvious which operation should be applied first.

However, our construction ensures that at any fixed integer time, the two-qubit gates form a bipartite graph, where space-like legs always connect a Z-spider (green) to a X-spider (red). As a result, the ZX-calculus fusion rules [Eqn.~\ref{eq:fusion}] guarantee that the gates commute, and the order of application does not affect the overall circuit. This is illustrated in the following identities:
\begin{equation}
\raisebox{-0.4\height}{\includegraphics[width=0.5\textwidth]{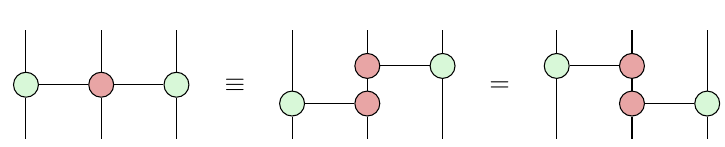}}
\end{equation}
\begin{equation}
\raisebox{-0.4\height}{\includegraphics[width=0.5\textwidth]{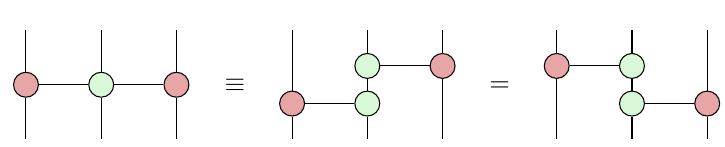}}
\end{equation}
These commutation relations continue to hold even when Hadamard gates are present on the legs, as long as the underlying bipartite structure of the spiders is preserved.

\subsection{Disorder-free limit: $p = 0$}

We begin by introducing the basic building block of the isotropic hyperdiamond circuit. In ($d$+1)-dimensional spacetime, each building block corresponds to a spider with $d$+2 legs, arranged such that the angle between any pair of legs is equal. For instance, the honeycomb lattice in $(1+1)$d consists of spiders with three legs, while the diamond lattice in $(2+1)$d has spiders with four legs.

There are two types of building blocks: one with a Z-spider at the center and one with an X-spider. For spiders of the same type, the legs point outward in the same directions. Let us define the unit vectors representing the directions of the legs extending from a Z-spider as $\hat{v}_1, \cdots, \hat{v}_{d+2}$. These vectors satisfy the isotropy condition:
\begin{align}
    \forall i\neq j, \quad \hat{v}_i \cdot \hat{v}_j = -\frac{1}{d+1}
\end{align}

Among the $d$+2 legs, we designate first two legs ($\hat{v}_1$ and $\hat{v}_2$) as the time-like legs, while the remaining $d$ legs are space-like. The time direction is defined as being proportional to $\hat{v}_1 - \hat{v}_2$, and all space-like legs are orthogonal to this direction:
\begin{align}
    \forall i \in \{3,.\cdots, d+2\},\quad (\hat{v}_1 - \hat{v}_2) \cdot \hat{v}_i = 0.
\end{align}

We also define a complementary building block with an X-spider at the center. Its legs point in the opposite directions: $\hat{v}_i' = -\hat{v}_i$. Similarly, the time-like legs are $\hat{v}_1'$ and $\hat{v}_2'$, and other $d$ legs are space-like.
These two building blocks are illustrated below for $(1+1)$d, $(2+1)$d, and $(3+1)$d:
\begin{equation}\label{eq:ZX_building_block}
\raisebox{-0.4\height}{\includegraphics[width=0.9\textwidth]{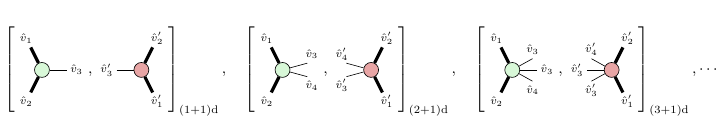}}
\end{equation}
where the thick legs are the time-like legs.

We construct the isotropic hyperdiamond lattice by connecting each Z-spider to an X-spider, pairing legs with matching indices: $\hat{v}_i$ of the Z-spider connects to $\hat{v}_i'$ of the X-spider. This defines a regular bipartite lattice, where the vertices alternate between Z- and X-spiders. Importantly, each space-like leg corresponds to a CNOT gate, as defined in Eqn.~\ref{eq:disorder_free}.

Using the ZX-calculus identity in Eqn.~\ref{eq:hadamard-rule}, each X-spider can be converted into a Z-spider, with Hadamard gates pushed out to its legs. Because the lattice is bipartite, this transformation yields a lattice composed entirely of Z-spiders, with all bulk edges decorated by Hadamard gates. At the boundary, the presence or absence of Hadamard gates may depend on the specific boundary conditions.

The explicit example of $d = 1$ is shown below. Starting from the original implementation with CNOT gates acting on qubits arranged on a Cartesian grid $\textbf{r}\in \mathbb{Z}^{d=1}$, the circuit maps onto a ZX-diagram with alternating spiders, and then to a representation consisting solely of Z-spiders with Hadamard-decorated edges. Although this construction is initially defined on the Cartesian grid, it is equivalent to a honeycomb lattice after an appropriate lattice deformation.
\begin{equation}
\raisebox{-0.4\height}{\includegraphics[width=0.9\textwidth]{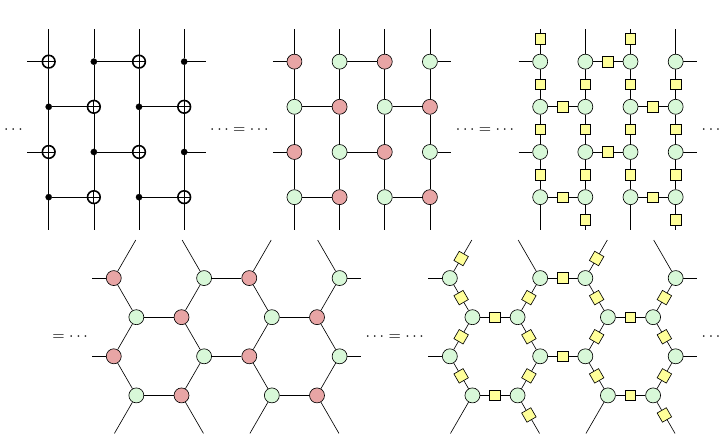}}
\end{equation}
The resulting honeycomb lattice consists of the building blocks of Z- and X-spiders introduced in Eqn.~\ref{eq:ZX_building_block}.

\subsection{Including disorder: $p>0$}
We previously explained how CNOT gates applied to qubits arranged on a Cartesian grid can be represented by an isotropic hyperdiamond lattice, with Z-spiders as the vertices and Hadamard-decorated edges. We now describe how to introduce randomness into this circuit in a controlled way.

The basic idea is to erase the Hadamard gates on the edges with probability $p$. For the time-like legs, this is accomplished by simply applying an additional Hadamard gate, since $\hat{H}^2 = \hat{\mathbb{I}}$. This is why we include the single-qubit gate $G_1(t, \textbf{r})$, which acts between the two-qubit layers at times $t$ and $t+1$.

For space-like legs, we erase the Hadamard gate by replacing the CNOT gate (which connects a Z-spider and an X-spider) with a diagram containing an intermediate Hadamard gate:
\begin{equation}\label{eq:zxmeas}
\raisebox{-0.4\height}{\includegraphics[width=0.5\textwidth]{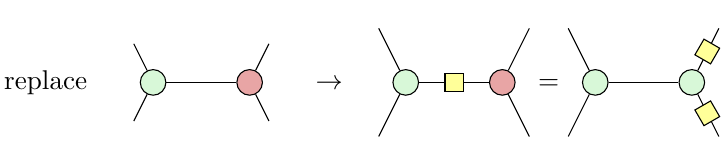}}
\end{equation}
In the circuit implementation, this is realized by performing a projective measurement of $\hat{Z} \hat{X}$. Although the measurement yields a random outcome of states, the resulting stabilizer group is the same up to sign, and thus the entanglement structure is equivalent. This defines the two-qubit gate $G_2(t, \textbf{r}_C, \textbf{r}_T)$: it acts as a CNOT with probability $1 - p$ and performs a $\hat{Z}(\textbf{r}_C)\hat{X}(\textbf{r}_T)$ measurement with probability $p$.

\subsection{Motivation for using the hyperdiamond architecture}
As briefly discussed in the main text, the hyperdiamond lattice is particularly well-suited for capturing the physics of interest. The first condition we required is that entanglement growth should not be suppressed in the $p = 0$ limit. In certain other lattice structures, such as simple hypercubic models, the butterfly velocity $v_B$ diverges in the STTI limit, which pathologically suppresses the entanglement growth, which we want to avoid. In contrast, our model remains fully unitary in the STTI limit, consisting entirely of CNOT gates. As a result, the entanglement grows generically for arbitrary initial states, except for special cases like Z-polarized product states.

The second condition is that the entanglement membrane should be pinned at $x = 0$ to clearly reveal a cusp. In our hyperdiamond lattice, one of the space-like legs, say, $\hat{v}_3$, is aligned with the direction normal to the entanglement membrane of interest, denoted by $\hat{x}$ in the main text. Consequently, the minimal free-energy membrane configuration is pinned at $x = 0$, bisecting all the $\hat{v}_3$ legs (connected to $\hat{v}'_3$) at that location. When a tilt is introduced, the membrane deviates from this pinned orientation and necessarily acquires additional steps, which cost extra free energy. This leads to the cusp (discontinuity in $\partial S/\partial x$) in the entanglement profile $S(x, t)$ at $x = 0$ in the smooth phase.

\section{Scaling relation of the average free energy with tilt}
As explained in the main text, we are interested in the growth of entanglement, with a focus on the bipartite entanglement entropy as our key observable. This quantity naturally probes the free energy of the entanglement membrane, rather than directly capturing its geometric displacement. In particular, we access the membrane's free energy response to deviations from the lattice-pinned configuration by varying the bipartition location $x$, which corresponds to tilting the membrane away from its pinned orientation $x=0$. In this section, we develop the scaling theory for this setup.

We consider a three-dimensional entanglement membrane embedded in (3+1)-dimensional spacetime. The membrane is anchored at $x = 0$ and $t = 0$, and we examine its free energy at later times $t$ and bipartition locations $x$. Both the initial and final anchor regions have area $L^2$. We define the disorder-averaged free energy at time $t$ and position $x$ as
\begin{equation}
 S(x,t,p) \equiv \mathbb{E}_{\eta} [ S(x,t,p,\eta)]~,
\end{equation}
where $\eta$ denotes a specific realization of disorder (in our case, the pattern of Hadamard-erased edges), and $p$ is the disorder parameter of the circuit. We fix the circuit runtime to scale with the system size, $t = rL$, where $r$ is an $O(1)$ constant. While the precise form of the scaling function may depend on $r$, the scaling relations and critical exponents do not. We present the scaling arguments with $t = 2L$, to match the numerical results.
 
The entanglement membrane is pinned at $x=0$ at time $t=0$, and at time $t$ it is pinned at $x$.  It is free to find its lowest $S$ configuration between time $0$ and $t$, with these constraints at the initial and final times.  So far, we only consider integer-valued $x$.  The entropy then follows
\begin{align}
    \delta S(x,t,p) \equiv S(x,t,p) - t^3s_{a}(p) - t^2 s_b(p) \approx t^{\theta_c}F\left(x,(p-p_c)t^{{1/\nu}}\right) ~,
\end{align}
where $s_a(p)$ and $s_b(p)$ are analytic backgrounds that are non-singular through the transition at $p_c$: $s_a(p)$ is the bulk contribution, $t\times(t/2)\times(t/2)\sim t^3$ is the membrane volume.  The membrane is pinned at upper and lower edges over an area of $L^2=(t/2)^2\sim t^2$, and this contributes to the analytic background $s_b(p)$.   
$F(x,y)$ is the scaling function, where we define $y\equiv(p-p_c)t^{1/\nu}$, and $\nu$ the correlation length exponent where the correlation length $\xi\sim|p-p_c|^{-\nu}$. Hence $y\sim (t/\xi)^{1/\nu}$.  The scaling function $F$ has a third argument, namely $r$, but here we set $r=2$ and do not display this argument.
The roughening transition critical point is known to be logarithmic rough \cite{Emig_Nattermann_1998_FRGprl,Emig_Nattermann_1998_FRGlong,Noh_Rieger_2002_numerics}, with the roughening exponent $\zeta=0$ with no characteristic lengths; therefore, the first argument in $F(x,y)$ is $x/t^\zeta=x/t^0=x$. To leading order in the $\epsilon = 4-d$ expansion~\cite{Emig_Nattermann_1998_FRGprl}, 
\begin{equation}\label{eq:thetac}
    \theta_c = 2 - \frac{\epsilon}{2} +\cdots = \frac{3}{2}+\cdots \quad (d=3)~.
\end{equation}
The scaling function has various regimes, as we consider and explore below: 

\subsection{At the critical point $y=0$ ($p=p_c$)}
At the critical point $y=0$ with no imposed tilt, $x=0$, we have $\delta S \approx t^{\theta_c}F(0,0)$.  Why does $\delta S$ depend on $t$?  In the limit of large $t$, the free energy density of the membrane at $p_c$ is $s_a(p_c)$.  This has contributions from negative entanglement entropy steps that come in at all scales.  For finite $t$, these negative entropy contributions from the largest scales beyond $t$ are not there.
Thus, this argument/scenario suggests that we should expect that $F(0,0)>0$ to be positive and of order one.

In the intermediate range of small tilts $t\gg|x|\gg 1$ we expect that at the critical point the scaling function behaves as
\begin{align}
    0<F(x,0) \sim |x|^c\quad(1<c<2)~.
\end{align}
Note that at small $|x|$, we expect $S(x)-S(0) \sim |x|^1$ for $y<0$ where the system at $x=0$ is in the smooth phase, and $S(x)-S(0)\sim x^2$ for $y>0$ where the system at $x=0$ is in the rough phase, so we expect this exponent $c$ at the critical point to be intermediate between these two behaviors.

Our numerical results with crossings of step ratios $\Delta S(|x|=1)/\Delta S(|x|=2)$ is telling us about the behavior of the differences between $F(0,0)$, $F(1,0)$, and $F(2,0)$, which should be universal, so the crossing should be a good indicator of $y=0$. 

In the limit of $x\sim t$, $\delta S$ should be proportional to its deviated volume $\delta S(x\sim t,t,p_c)\simeq t^{\theta_c}t^c\sim t^3$, so $c=3-\theta_c$. Therefore, for $1\ll |x|\ll t$
\begin{align}
    \label{eqn:critical_F}
    F(x,0) &\sim |x|^{3-\theta_c}\quad(\Rightarrow 1<\theta_c<2)~,\\
    \delta S &\sim t^{\theta_c} |x|^{3-\theta_c} = t^3 \left(\frac{|x|}{t}\right)^{3-\theta_c}~.
\end{align}
This is the leading scaling behavior in this regime of $1\ll |x|\ll t$, and we expect that this behavior is self-averaging, so the variations between realizations of the randomness are subleading to this.  However, due to $|x|\gg 1$, in this regime, the membrane is not aligned with the lattice, and it becomes rough due to the randomness.  The length scale beyond which the system ``knows'' it is tilted is $\xi_t\sim t/|x|$.  On scales below that, it is critical, while beyond that, it is rough.  Thus the pinning (and non-self-averaging) contribution should scale as $\sim\xi_t^{\theta_c}(t/\xi_t)^{\theta_r}\sim t^{\theta_c}|x|^{\theta_r-\theta_c}$, which is indeed subleading in this regime of $\delta S$.
This is by rough analogy to the DPRE in (1+1)d, where the $\sim L^1$ term in the energy is the same for all samples (so self-averages), but the $\sim L^{1/3}$ term varies between samples so both its mean and its standard deviation are $\sim L^{1/3}$ (non-self-averaging).

\subsection{In the smooth phase $y<0$ ($p<p_c$)}
\subsubsection{$x=0$}
When $x=0$, the scaling function $F(0,y)$ is a function of a single parameter $t/\xi\sim y^\nu$.  For $t\gg\xi$ ($|y|\gg 1$)
\begin{align}
\label{eqn:smooth_x=0}
    0<\delta S \sim \xi^{\theta_c}\left(\frac{t}{\xi}\right)^3 = t^3\xi^{\theta_c - 3}\sim t^3|p-p_c|^{(3-\theta_c)\nu}\sim t^{\theta_c}|y|^{(3-\theta_c)\nu}~,
\end{align}
which is partitioning the volume into patches of length scale $\xi$ and each patch has a contribution of $\sim\xi^{\theta_c}$.  At $p_c$ the system puts in negative energy steps at all scales.  In the smooth phase, it cannot put in negative energy steps at large scales beyond $\xi$ because there are no such steps to put in, so this causes a singular increase in $\delta S$ relative to $p_c$.  This is a $\sim t^3$ term like the analytical background $s_a(p)$, but when we take $t\rightarrow\infty$ first before taking $p\rightarrow p_c$ it is singular in $(p-p_c)$.
It is positive, so we have $0<F(0,y)\sim |y|^{(3-\theta_c)\nu}$ for $y<0$ and $|y|\gg 1$ and this matches on smoothly to an order-one positive $F(0,0)$ at $y=0$.

\subsubsection{small $|x|$}
The partitioning location $x$ is always an integer (with a unit of the lattice constant), so we have a natural UV cutoff and need not worry about the limit $ x\rightarrow0$.
If we are in the smooth phase, the membrane has $|x|$ number of extra steps for small $|x|$ (we will discuss the meaning of $|x|$ being small and describe this crossover in the following subsection) and each step has an area of $(t/2)^2$. Therefore, $S(x,t) - S(0,t) \simeq f_{\mathrm{step}}|x|t^2/4$, where $f_{\mathrm{step}}$ is the step free energy per unit area. This leads us to think that the scaling function should satisfy
\begin{align}
\label{eqn:smooth_smallx}
    0<F(x, y) - F(0,y) \sim |x| |y|^{(2-\theta_c)\nu}\quad(\mathrm{small~enough~}|x|>0,~y\ll -1)~,
\end{align}
and $f_{\mathrm{step}}$ scales as
\begin{align}
    f_{step} \sim|p-p_c|^{(2-\theta_c)\nu}.
\end{align}
From Eqn.~\ref{eqn:smooth_x=0} and Eqn.~\ref{eqn:smooth_smallx}, we derive the scaling function and the entanglement entropy should follow
\begin{align}
    \label{eqn:smooth_F}
    F(x,y) &\cong A|y|^{(3-\theta_c)\nu} + B|x||y|^{(2-\theta_c)\nu}\quad(\mathrm{small~enough~}|x|\mathrm{~and~} y\ll -1)~,\\
    \delta S(x,t,p)&\cong At^3 (p_c-p)^{(3-\theta_c)\nu} + B|x|t^2(p_c-p)^{(2-\theta_c)\nu}~,
\end{align}
with $A>0$ and $B>0$.

\subsubsection{crossover to the tilted regime}
The membrane we consider has the boundary condition that it is pinned at $(x,t)$ and $(0,0)$. Therefore, the characteristic temporal length scale to notice the tilt (by one unit) is $\xi_t\sim t/|x|$.\footnote{We stress that $t$ and $x$ are just the numbers with a unit of lattice constant.} The aforementioned scaling of ``small enough'' $|x|$ is when $\xi_t \gg \xi$, and we can identify where the crossover from the smooth phase to the tilted regime occurs:
\begin{align}
\begin{cases}
    \xi_t\gg\xi&\Leftrightarrow~~ t/|x|\gg\xi\quad\mathrm{smooth~phase~with~steps}~,\\
    \xi_t\sim\xi&\Leftrightarrow~~ t/|x|\sim\xi\quad\mathrm{crossover}~,\\
    \xi_t\ll\xi&\Leftrightarrow~~ t/|x|\ll\xi\quad\mathrm{tilted~regime}~.
\end{cases}
\end{align}
At the crossover, the two length scales $\xi_t\sim\xi$ are comparable, and $|x|\sim t/\xi \sim |y|^\nu$. In the crossover regime, Eqn.~\ref{eqn:smooth_F} matches Eqn.~\ref{eqn:critical_F}. This suggests a scaling function to be reduced with a function $G$ described by a single parameter $\xi/\xi_t$:
\begin{align}
    F(x,y) \sim |x|^{3-\theta_c}G\left(\frac{y}{|x|^{1/\nu}}\right)~,
\end{align}
where $G$ is smooth and positive near zero, and diverges appropriately at large argument to match the above (and below) forms.  In particular, $0<G(z)\sim |z|^{(3-\theta_c)\nu}$ for large negative argument $z$.  This reduction appears to apply in the smooth phase and the critical regime when either $|x|\gg 1$ or $|y|\gg 1$ or both.  It does not apply when both $x$ and $y$ are of order one or less.  In this latter regime $F(x,y)$ is a smooth order-one function of $y$ for each order-one (or zero) integer $x$. 

\subsection{In the rough phase $y>0$ ($p>p_c$)}
The rough phase is characterized with two exponents $\zeta_r$ and $\theta_r$, where considering a membrane of a characteristic length $L$, the variation of heights scales as $\sim L^{\zeta_r}$ and the subleading contribution to the membrane free energy scales as $\sim L^{\theta_r}$. The two exponents satisfy the hyperscaling relation (termed ``tilt symmetry'' in the context of elastic manifolds~\cite{Emig_Nattermann_1998_FRGprl}) $\theta_r = d-2 + 2\zeta_r = 1 + 2\zeta_r$ in (3+1)d. FRG estimates to lowest order in $\epsilon$ of the exponents are \cite{Emig_Nattermann_1998_FRGprl,Emig_Nattermann_1998_FRGlong}
\begin{equation}\label{eq:FRG}
\zeta_r \cong 0.21 \rightarrow \theta_r \cong 1.42~.
\end{equation}

\subsubsection{$x=0$}
Deep in the rough phase, where $t\gg\xi$ and thus $y\gg 1$, the free energy fluctuates with the rough phase exponent $\theta_r$, and the mean of this fluctuation is nonzero and follows $\delta S \sim t^{\theta_r}$.  Relative to the logarithmically-rough critical state, the system at large scales is rougher, to lower its free energy by finding the lowest-energy positions. This suggests that $\delta S<0$ in this regime.
Therefore, the scaling function and mean free energy scales
\begin{align}
\label{eqn:rough_x=0}
    0>F(0,y)&\sim |y|^{(\theta_r - \theta_c)\nu}~,\\
    \delta S &\sim \xi^{\theta_c}\left(\frac{t}{\xi}\right)^{\theta_r}~.
\end{align}
As $y$ is reduced to zero, this must match smoothly to the positive $F(0,0)$, so, if $F(y,0)$ indeed goes negative, this scaling function must change sign at some $y$ that is positive and of order one, where $\xi\sim t$.  Note that the exponents $\theta_c$ and $\theta_r$ are not very different, so this scaling function $F(0,y)$ has a rather weak dependence on $y$ for $y\gg 1$.

\subsubsection{small $|x|$}
Deep in the rough phase, the step free energy vanishes ($f_{\mathrm{step}}=0$), and the membranes ignore the lattice structure on long length scales $\gg\xi$.  It is at and below the length scale $\xi$ where the system is critical and being affected by the lattice.  If we weakly tilt, the tilt at length scale $\xi$ is $|x|_\xi=|x|\xi/t= \xi/\xi_t \ll 1$.  In this regime the free energy increase to tilt each volume $\sim\xi^3$ patch should be a smooth even function of $x_\xi$, which we expect to be $\delta S_\xi\sim\xi^{\theta_c}x_\xi^2\sim \xi^{2+\theta_c} x^2 / t^2$. 
There are $\sim(t/\xi)^3$ total patches, so
\begin{equation}\label{eq:S-tilt-rough}
S(x,t) - S(0,t) \sim \xi^{\theta_c-1} x^2 t.
\end{equation}
The scaling with $x$ and $t$ matches with the intuition that, at length scales $t \gg \xi$, we can just ignore the lattice and consider the extra volume to be $(t/2)^2 (\sqrt{t^2 + x^2} - t) \sim x^2 t$. To match this dependence on $x$ and $t$, the scaling function should be
\begin{align}
\label{eqn:rough_smallx}
    F(x, y) - F(0,y) \sim x^2 y^{(1-\theta_c)\nu}\quad(\mathrm{small~}|x|,~y\gg1)~.
\end{align}
which reproduces the $\xi^{\theta_c-1}$ dependence predicted in Eqn.~\ref{eq:S-tilt-rough}.

From Eqn.~\ref{eqn:rough_x=0} and Eqn.~\ref{eqn:rough_smallx}, we derive the scaling function and the free energy
\begin{align}
    \label{eqn:rough_F}
    F(x,y) &\cong A'y^{(\theta_r-\theta_c)\nu} + B'x^2y^{(1-\theta_c)\nu}\quad(\mathrm{small~}|x|\mathrm{~and~} y\gg1)~,\\
    \delta S(x,t,p)&\cong A't^{\theta_r}(p-p_c)^{(\theta_r - \theta_c)\nu} + B'tx^2(p-p_c)^{(1-\theta_c)\nu}~.
\end{align}
with $A'<0$ and $B'>0$ each of order one.  These two terms become comparable when $x\sim(t/\xi)^{\zeta_r}$ which matches the roughness of the membrane in this regime.  But there does not seem to be any reason to expect another crossover at this tilt; for larger tilts, the tilt term becomes dominant, but should remain of this form.

\subsubsection{crossover to the tilted regime}
As in the smooth phase, we use the same criteria to locate the crossover between the weakly and strongly tilted regimes by comparing $\xi_t = t/x$ and $\xi\sim |p-p_c|^{-\nu}$.  Note this is all within the regime of tilts where $|x|\ll t$.  At the crossover when $\xi_t\approx\xi\Leftrightarrow x\approx t/\xi\sim y^\nu$, Eqn.~\ref{eqn:rough_F} becomes
\begin{align}
    F(x,y)\cong A' x^{\theta_r - \theta_c} + B' x^{3-\theta_c}\quad(\mathrm{crossover)}~.
\end{align}
Since the first term is negligible compared to the second term ($\theta_r - \theta_c < 3-\theta_c$), our crossover criteria match two regimes of Eqn.~\ref{eqn:critical_F} and Eqn.~\ref{eqn:rough_F}.

\section{Fluctuations of the free energy with tilt}
In the previous section, we considered the disorder-averaged free energy of the entanglement membrane. Here, we turn to the fluctuations of the free energy across disorder realizations.
First consider a membrane without any tilt. In the smooth phase, the membrane spans a volume of $O(t^3)$, collecting $\sim t^3$ independent local contributions to the free energy (we are keeping $l\sim t$). By the central limit theorem, the standard deviation over disorder realizations of the total free energy then scales as $\sim t^{3/2}$. In contrast, in the rough phase, the membrane configuration is determined by a nontrivial minimization of free energy over all configurations it can take, and the fluctuations between disorder realizations follow an extreme value distribution. In this case, the standard deviation scales as $\sim t^{\theta_r}$, where $\theta_r$ is the roughening exponent. At the roughening critical point, we expect the fluctuations to scale as $\sim t^{\theta_c}$. However, since the values of $\theta_c$, $\theta_r$, and $3/2$ are close for a three-dimensional membrane, numerically distinguishing these regimes is likely to be challenging.

Instead, we propose a more sensitive diagnostic: the standard deviation over disorder realizations of the \textit{difference} within each single disorder realization in free energy between two nearby-tilted entanglement membranes:
\begin{align}
\sigma[\Delta S(\pm1)] \equiv \sigma_\eta [S(x=1,t,p,\eta)-S(x=0,t,p,\eta))]~,
\end{align}  
where $\sigma_\eta$ is a standard deviation over the disorder realizations $\eta$.
In the smooth phase, the two entanglement membranes, corresponding to $S(x=0)$ and $S(x=1)$, typically differ over a region of volume $\sim t^3$, where the membrane for $S(x=1)$ is shifted by one unit in the $x$ direction. This leads to a standard deviation of order $\sim t^{3/2}$.  The mean difference is the step free energy $\sim t^2$, so the step free energy self-averages.
In the rough phase, however, a change in tilt by one unit is small compared to the typical roughness $\sim t^{\zeta_r}$. The two membranes thus typically coincide over most of the bulk and differ only near the top edge, where they are pinned differently over this area $\sim t^2$. Therefore, the corresponding fluctuations scale as $\sim t$. Notably, the mean difference is also of order $\sim t$, so this difference does not self-average in the rough phase.

Additionally, there are potentially strong sub-leading contributions to the fluctuations in the smooth phase. For $S(x=1)$---which corresponds to a tilted membrane---the membrane has a step. This step can be viewed as a two-dimensional interface between three-dimensional domains of heights 0 and 1 of the membrane.  Viewing these two heights as the analogs of spin up and down, the step maps to an interface in a three-dimensional \textit{random field} Ising model.  In such models, the interface (in this case, the step) is rough, with typical displacements (parallel to the membrane) scaling as $\sim L^{2/3}$~\cite{Peled2504}, as predicted by an Imry-Ma argument based on comparing elastic and disorder contributions~\cite{Imry_Ma_prl1975,grinstein_ma_1983,villain_prl1984}. 
This sub-leading effect contributes free energy fluctuations of order $\sim L^{4/3}\sim t^{4/3}$, which is not much smaller than the leading $\sim t^{3/2}$ fluctuations from the bulk of the membrane.

\end{document}